\pdfoutput=1
\documentclass[aps,prb,twocolumn,superscriptaddress]{revtex4}

\usepackage{graphicx}

\begin{document}

\title{Electrochemical ferroelectric switching: The origin of polarization reversal in ultrathin films}

\author{N. C. Bristowe}
\affiliation{Theory of Condensed Matter,
             Cavendish Laboratory, University of Cambridge, 
             Cambridge CB3 0HE, UK}
\affiliation{Department of Earth Sciences, University of Cambridge, 
             Downing Street, Cambridge CB2 3EQ, UK}
\author{Massimiliano Stengel}
\affiliation{ICREA - Instituci\'o Catalana de Recerca i Estudis Avan\c{c}ats, 08010 Barcelona, Spain}
\affiliation{Institut de Ci\`{e}ncia de Materials de Barcelona (ICMAB-CSIC), Campus UAB, 08193 Bellaterra, Spain}
\author{P. B. Littlewood}
\affiliation{Theory of Condensed Matter,
             Cavendish Laboratory, University of Cambridge, 
             Cambridge CB3 0HE, UK}         
\affiliation{Physical Sciences and Engineering,
             Argonne National Laboratory, 
             Argonne, Illinois 60439, USA}    
\author{J. M. Pruneda}
\affiliation{Centre d'Investigaci\'{o}n en Nanoci\'{e}ncia i Nanotecnologia (CSIC-ICN), Campus UAB, 08193 Bellaterra, Spain}             
\author{Emilio Artacho}
\affiliation{Department of Earth Sciences, University of Cambridge, 
             Downing Street, Cambridge CB2 3EQ, UK}

\date{\today}

\begin{abstract}

Against expectations, robust switchable ferroelectricity 
has been recently observed
in ultrathin (1 nm) ferroelectric films exposed to air [V. Garcia $et$ $al.$,
Nature (London) {\bf 460}, 81 (2009)].
  Based on first-principles calculations, we show that
the system does not polarize unless charged defects or adsorbates form
at the surface. We propose electrochemical processes as the most likely 
origin of this charge. 
  The ferroelectric polarization
of the film adapts to the 
external ionic charge generated on its surface
by redox processes when poling the film.
This, in turn, alters the band alignment at the bottom electrode interface,
explaining the observed tunneling electroresistance.
Our conclusions are supported by energetics calculated for varied
electrochemical scenarios.  
\end{abstract}

\pacs{}

\maketitle




\section{Introduction}

  Complex oxides have long been viewed as possible
candidates for the next generation of electronic 
devices, which require reduced feature sizes, enhanced operating speeds
and low consumption.
  Amongst oxides, ferroelectrics offer
the ability to store information in a non-volatile manner 
via fast reversible polarization switching in ferroelectric random-access
memory (FeRAM).
  The observation of giant tunneling electroresistance (TER)~\cite{Garcia2009} 
in ultrathin (3 unit cells) ferroelectric films has recently opened a novel
paradigm for device design based on these materials~\cite{Zubko2009,Segal2009}.

Although the experiments~\cite{Garcia2009} ascribed TER 
to ferroelectricity, which appeared robust and switchable, how
the polar state is stabilized in such thin films is by no
means established.
In principle, a ferroelectric film with an exposed surface cannot sustain a 
monodomain polarization perpendicular to the surface, because of
the strong depolarizing field that would inevitably arise~\footnote{
%
The polarization is clearly observed to be perpendicular to the interface, consistent
with the expected behavior of compressively strained films~\cite{Schlom2007}.
}.
Charged particles from the environment could in principle  
cancel the depolarizing field~\cite{Fridkin1980} (Fig. 1 left).
So far, however, the only chemical control of switching in air relates to neutral 
species, O$_2$~\cite{Wang2009,Fong2006,Spanier2006} (Fig. 1 center).
It is then not clear how neutral gas-phase molecules could interact with a 
biased atomic force microscopy (AFM) tip to produce the polar state.

Here we argue that the voltage applied with the AFM tip 
induces electrochemical switching (Fig. 1 right), i.e. redox processes that are essential to 
liberate free charge and therefore screen the depolarizing field.
This process would act as a nanobattery, rather than a nanocapacitor. 
Note that the same mechanism could explain other effects at oxide interfaces,
such as the switchable two-dimensional electron gas (2DEG) at the LaAlO$_3$/SrTiO$_3$ interface~\cite{Cen2008,Bristowe2011a},
where the switching appears to be mediated by surface charge~\cite{Xie2010}.


  
\begin{figure}[b]
\includegraphics[width=0.45\textwidth]{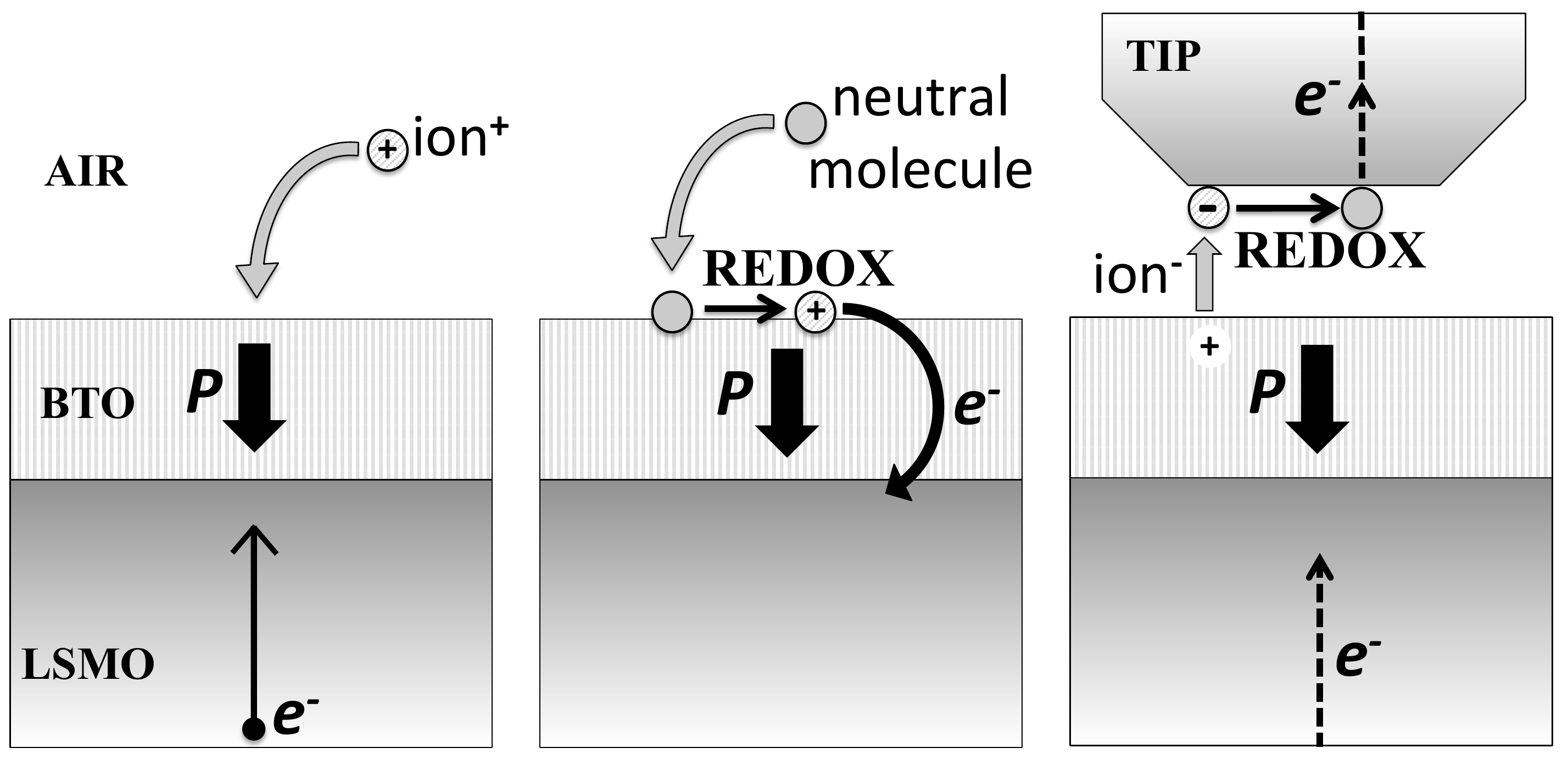}
\caption{\label{BANDS}{
Schematic illustration of the conventional (left) and redox 
(center) mechanisms for ferroelectric screening in the absence of a top
electrode. The presence of a biased tip can promote an alternative
redox mechanism that provides an 
external circuit for the screening electrons (right).  
}}
\end{figure}

  To explore this mechanism we consider the system 
studied experimentally in Ref.~\onlinecite{Garcia2009}, consisting of 
a compressively strained nanometer-thick BaTiO$_3$ (BTO) film
on a La$_{0.7}$Sr$_{0.3}$MnO$_3$ (LSMO) bottom electrode.
  Here we show, using first principles calculations, 
  that 
(i) the pristine system (clean BTO surface with an ideal TiO$_2$ termination) 
   does not allow for a ferroelectric polarization, $P$, normal to the surface despite the
   large compressive strain; (ii) a 
   non-zero $P$ 
   is crucially dependent on the presence of a surface 
   external ionic charge,
   in the form of defects
   or adsorbates;  and (iii) 
the energetics for the formation of
oxidized or reduced surface defects support the
electrochemical switching model.
  We also find (iv) a systematic change in band offset with screening 
  charge density, which we identify as the microscopic mechanism behind 
  the experimentally observed TER~\cite{Garcia2009},
%
  and (v) a large magnetoelectric coupling, 
due to the accumulation or depletion of spin-polarized carriers at
the interface with ferromagnetic LSMO.
  The connection between these effects can be summarized as follows:
under open-circuit boundary conditions
the electric displacement field $D$ within the film, 
the change in magnetization at the interface $\Delta M$
and the interface dipole, 
are all proportional (or equal) to the external ionic charge density,
$Q$ per unit surface $S$, produced by the redox processes.  

\section{Methods}


  The  density-functional theory (DFT) calculations are performed using the 
spin-polarized Wu-Cohen (WC) exchange-correlation functional~\cite{Wu2006a}, 
as implemented in the {\sc Siesta} code~\cite{Ordejon1996,Soler2002}~\footnote{
  Details of the pseudopotentials, numerical atomic orbitals and LSMO doping are given in Ref.~\onlinecite{Ferrari2006} and~\onlinecite{Junquera2003}.}.
  We find GGA-WC to reproduce bulk~\cite{Ferrari2006} and 
surface~\cite{Ferrari2006,Pruneda2007} properties of LSMO that were calculated
using the Perdew-Burke-Ernzerhof (PBE) scheme~\cite{Perdew1996}; at the same
time, GGA-WC is more appropriate for ferroelectric oxides.
  The LSMO/BTO system consists of 5.5 unit cells of LSMO 
(MnO$_2$-terminated) stacked with 3 unit cells of BTO along the 
$c$ direction in a slab geometry. 
The supercell contains a 15 \AA{} thick 
vacuum layer and has either 2$\times$2 or $\sqrt{2}$$\times$$\sqrt{2}$ 
in-plane periodicity (see Fig. 2).
  The 5.5 unit cells of LSMO are thick enough to show bulk-like features in the center,
and 3 unit cells of BTO was experimentally shown 
to be thick enough for ferroelectricity~\cite{Garcia2009}.
  We use a dipole correction to simulate open-circuit 
boundary conditions, enforcing zero macroscopic electric field  
in the vacuum layer.
%
  We constrain the in-plane lattice parameter to experimental bulk NdGaO$_3$ (NGO) to
reproduce the experimental conditions of Ref.~\cite{Garcia2009}; this 
imposes a large (3\%) compressive strain on BTO.
%
%
Based on this slab geometry, we perform a number of calculations where
we vary the surface composition by introducing defects or adsorbates. 
In particular, we simulate the clean TiO$_2$-terminated surface (we 
shall refer to this structure as ``pristine'' henceforth); one O vacancy 
(``O-vac'') or adatom (``O-ads'') per 2$\times$2 surface cell; one H 
adatom (``H'') or OH group (``OH'') per $\sqrt{2} \times \sqrt{2}$ cell
~\footnote{Atomic forces were relaxed to less than 40 meV/\AA.}. 
%
Hereafter we shall discuss
the results with special regard for the presence or absence of 
ferroelectric polarization in each case.

%

  
\begin{figure}[t]
\includegraphics[width=0.45\textwidth]{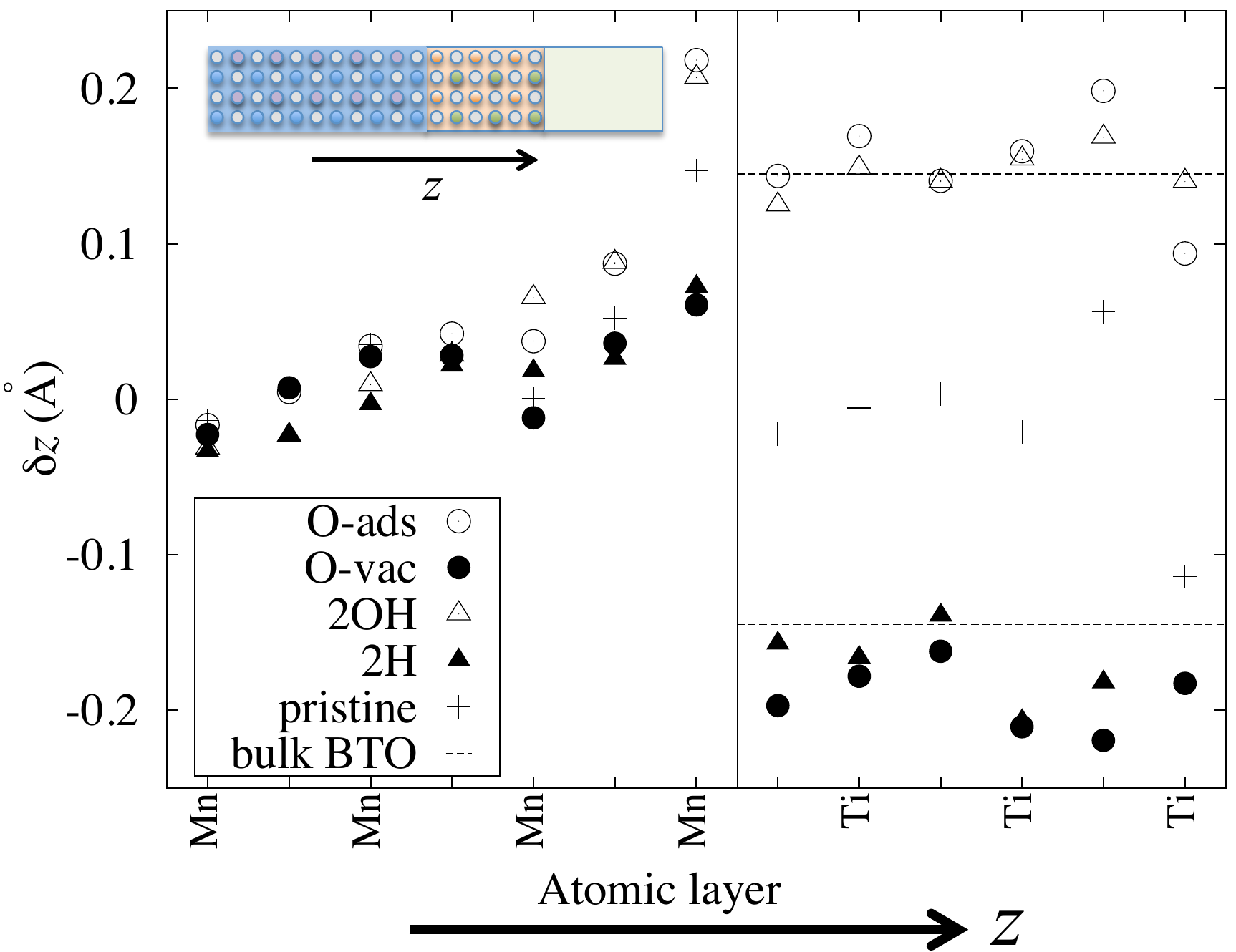}
\caption{\label{BAND}{(Color online) 
Cation-anion splittings, $\delta z=z_{\mathrm{cation}}-z_{\mathrm{anion}}$
through the LSMO/BTO slab (the bottom half of LSMO is not shown).
The dotted lines correspond to the average of the AO and BO$_2$ layer
anion-cation splitting for the inwards and outwards $P$ in bulk BTO, strained to NGO.
}}
\end{figure} 

  

\section{Discussion}

\subsection{The pristine system}

Fig. 2 shows the relaxed out-of-plane structural distortions as
a function of the surface chemical environment. 
The pristine system is characterized by negligible distortions
in the interior of the BTO film, suggesting the absence of 
macroscopic $P$ in this system. Only a surface rumpling is
present, resulting in a small net inwards dipole (non-switchable) 
that decays rapidly towards the bulk (a surface rumpling is a known 
general feature of oxide surfaces, in particular the TiO$_2$ 
termination of BTO~\cite{Padilla1997,Heifets1997}).
A vanishing $P$ is consistent with the open-circuit boundary
conditions, despite the large compressive strain.
 In absence of a top electrode the macroscopic electric 
displacement field $D$ in BTO is equal and opposite to the 
density of external surface charge. As this charge is zero at the
clean TiO$_2$ surface, the film is constrained to a paraelectric
state.

\subsection{Chemical switching}

  To illustrate possible screening scenarios, we now include representative 
surface defects~\footnote{
  Sampling the entire phase space (redox species and density, temperature,
partial pressure, polarization) is beyond the scope of this work, but 
can be done within thermodynamic theory (see Ref.~\onlinecite{Stephenson2011}).}.
The O-vac and O-ads systems are both characterized by
large ferroelectric distortions (Fig. 2). These
are comparable to the strained bulk, where we calculate a 
spontaneous polarization $P_0$=0.369 C/m$^2$ (0.35 $e/S$).
This result is again consistent with the constraint that $D=-Q/S$. 
In fact, one oxygen defect for every 2$\times$2 unit cells 
(0.5 $e/S$) yields a larger surface charge than what would be 
sufficient to screen $P_0$. This justifies the larger 
cation-anion rumplings that we obtain in the film compared with the 
bulk (Fig. 2).
%
OH and H adatoms (with $\sqrt2$$\times$$\sqrt2$ coverage 
to maintain $Q/S$) produce distortions of similar magnitude (Fig. 2).
%
This confirms the generality of the ferroelectric switching mechanism:
the ferroelectric state really depends on the net surface charge, and not on the chemical 
identity of the adsorbed species.


In order to study the electrochemical switching (Fig. 1 right), we commence 
by analyzing $chemical$ switching (Fig. 1 center). 
Both are controlled by redox processes that transform bound charge into free charge, allowing for an electronic transfer 
between the surface defect and the metal substrate,
but have different associated chemical sources/drains and energetics.
Chemical switching was recently shown in a system consisting of PbTiO$_3$ on SrRuO$_3$~\cite{Fong2006,Wang2009}
and BTO films on Au or vacuum~\cite{Spanier2006}.

To assess whether these redox reactions are thermodynamically accessible
in typical experimental conditions, we estimate the formation energy of 
the defective systems
taking the reactions: 1) Slab(pristine) $\rightarrow$ Slab(O-vac)+1/2O$_2$ 
and equivalent for O-ads, 2) 1/2H$_2$O+1/4O$_2$+Slab(pristine) 
$\rightarrow$ Slab(OH)  and 3) 1/2H$_2$O+Slab(pristine) $\rightarrow$ Slab(H)+1/4O$_2$.
The chemical potential of the relevant molecular species is set to the calculated total energy of the 
spin-polarized molecule in a large cubic box. 
%
%
%
The results are summarized in Table 1.
They suggests that, whilst the oxygen adatom is likely
to form under oxygen-rich conditions, the formation energy
for the oxygen vacancy is possibly too high to form even in oxygen-poor conditions.
%
The calculated OH and H formation energies
suggest that water is a very likely redox 
intermediate. 
%
Note that H$_2$O is ubiquitous in most experiments performed in air, 
and was recently found to play a crucial role in AFM experiments performed
on LaAlO$_3$/SrTiO$_3$~\cite{Bi2010}.
%
Since both sets of reactions involve oxygen, we therefore expect 
that altering the surrounding oxygen partial pressure would affect the stability 
of reduction or oxidation processes, consistent with the recently observed chemical 
switching~\cite{Wang2009}.
%

\begin{table}[b]
\caption{\label{multilayers}{The formation energy, $E_f$, of the defective systems 
for O$_2$ and H$_2$O rich conditions (see text for definitions).
}}
\begin{ruledtabular}
\begin{tabular}{ c | c c c c}
& O-vac & O-ads & OH & H \\ [3pt]
\hline 
$E_f$ (eV) & +3.6 & -0.4 & -1.5 & +0.9 \\
\end{tabular}
\end{ruledtabular}
\end{table}

\begin{figure}[t]
\includegraphics[width=0.42\textwidth]{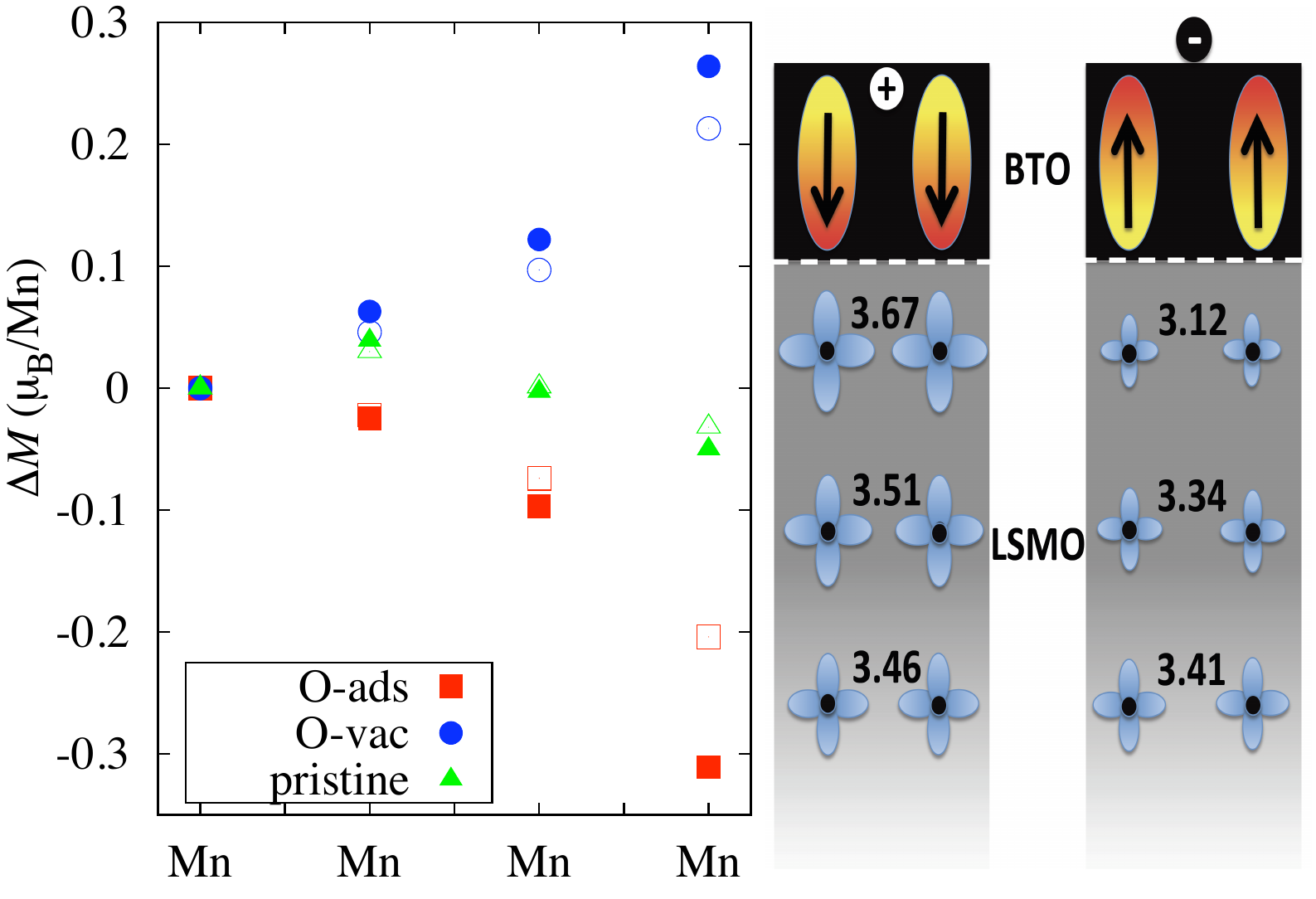}
\caption{\label{BAND}{(Color online) 
Change in Mn magnetic moment (left), $\Delta M$,
near the interface (the interface MnO$_2$ layer is on the right).
Open symbols represent the change in 3$d$ $e_g$ occupation, 
and closed symbols the total magnetic moment.
A schematic illustration (right) of the effect of O-vac and O-ads on the BTO polarization (arrows)
and the Mn  3$d$ $e_g$ occupation (blue lobes) and total Mn magnetic moment (numbers).
}}
\end{figure} 

\subsection{Electrochemical switching}

Now we discuss how the 
electrochemical processes could proceed in practice during the AFM switching 
experiments of Ref.~\onlinecite{Garcia2009} (general electrochemical processes 
on oxide surfaces are reviewed in Ref.~\onlinecite{Kalinin,Waser2009}).
%
As schematically shown in Fig. 1 (right), a biased tip close to contact can 
remove surface ions. These would then undergo a redox reaction at the tip surface.
This process is favored by the energy associated with the biased external 
circuit, $QV_{ext}$, but costs an energy equal to the change in binding 
energy of the ion to the ferroelectric surface and to the tip surface, $\Delta E_f$ 
(this effectively redefines the relevant chemical potential). 
By minimizing the Gibbs free energy of the system
(see e.g. Ref.~\onlinecite{Stephenson2011} or~\onlinecite{Morozovska2010}) 
it can be shown that poling can stabilize redox 
defects if $V_{ext}>\Delta E_f/Q$, after which the equilibrium redox charge 
density, $Q/S$, and polarization both grow with  $V_{ext}$. 
This electrochemical process would then act as a nanobattery,
rather than a nanocapacitor. 
By controlling the environment (species and chemical potential) 
and $V_{ext}$, one may be also able to selectively control the 
active redox reaction, potentially opening new routes to surface 
redox catalysis.
  After removal of the tip, the surface redox density from poling can remain, 
since the reverse reaction is now blocked by
key reactants being removed with the tip.
This would explain the observation of Ref.~\onlinecite{Garcia2009} that the
domains are stable for a very long time after ``writing''.
 Of course, lateral charge diffusion across domain boundaries~\cite{Kalinin2004} may 
still occur in principle, but kinetic barriers are likely to hinder such processes.
Therefore the bulk polarization, $P_0$, is expected to be an estimate of the equilibrium
polarization after poling. 
  We note that unlike in the LaAlO$_3$/SrTiO$_3$ system where the polarization is driving the 
surface chemistry~\cite{Bristowe2011a}, in ferroelectric films we expect it is the surface chemistry (and poling)
that is driving the polarization.
   This is because the energy scale for changing the polarization is much larger
 in LaAlO$_3$ than in the ferroelectric.


\begin{figure}[t]
\begin{center}
\includegraphics[width=0.4\textwidth]{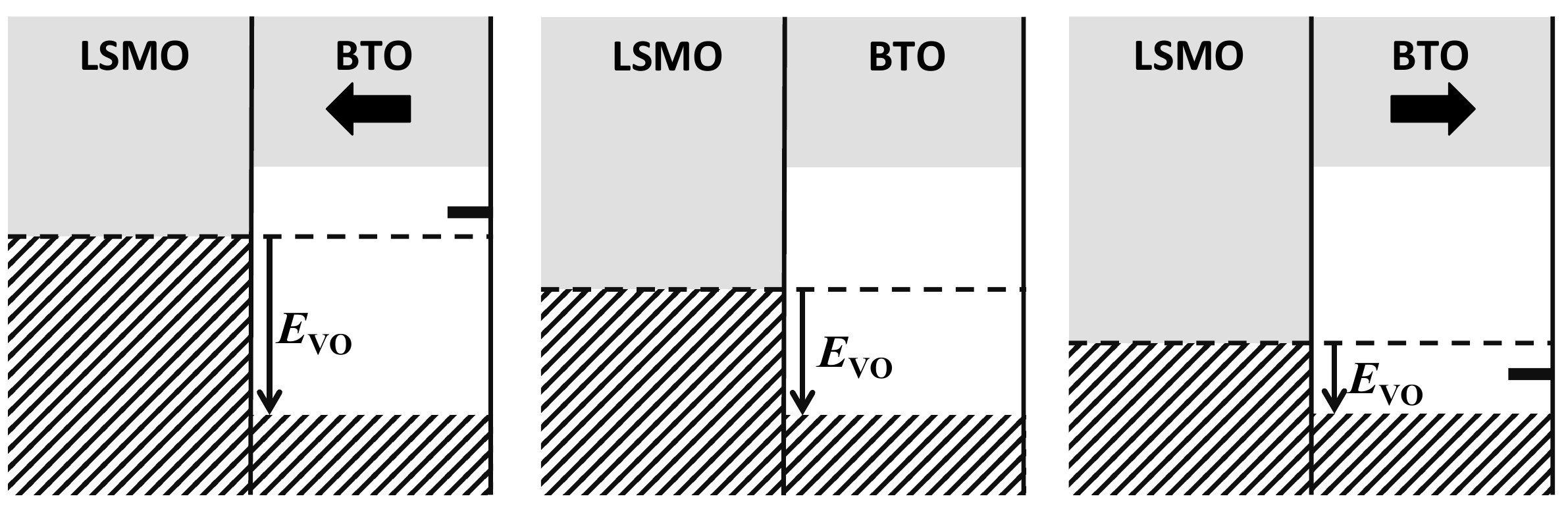}
\includegraphics[
width=0.4\textwidth,viewport= 50 5 800 620,clip]{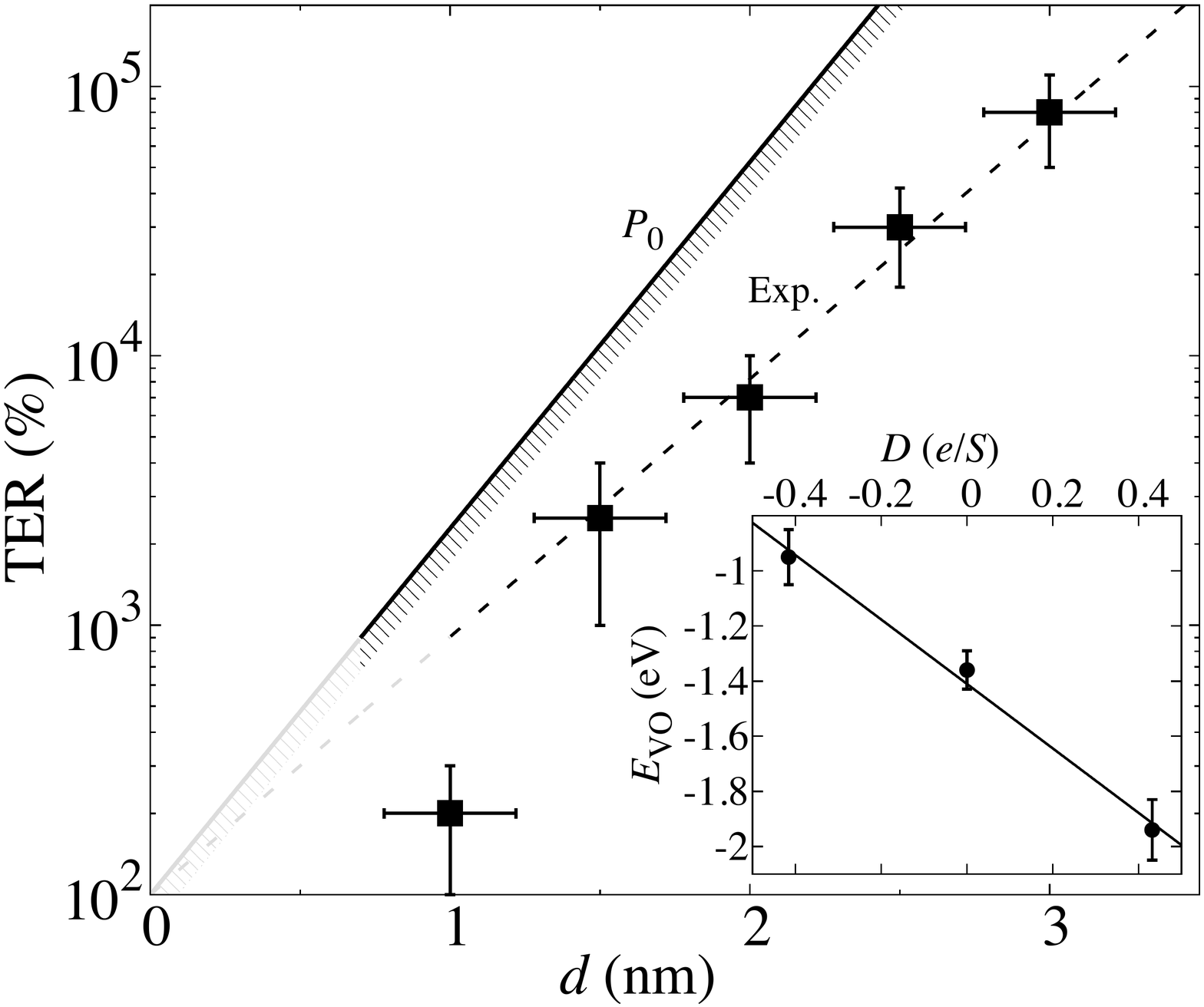}
\caption{\label{TRAP}{
Top: Schematic illustration of the change in
band offset, $E_{\mathrm{VO}}$, with polarization reversal.
Bottom: Tunnel electroresistance (TER)
vs BTO thickness. 
Experimental points taken from Garcia $et$ $al$.~\cite{Garcia2009}
(squares with dashed line fit) 
are compared with a theoretical expression~\cite{Gruverman2009} 
which uses the tunneling barrier height expected 
from the BTO bulk polarization, 
$P_0$ (solid line).
Inset: Calculated band offset against 
electric displacement field for the three BTO states.
The straight line fit is used to determine the band offset (and hence barrier height)
at $\pm P_0$ for the TER plot.
}}
\end{center}
\end{figure}

\subsection{Magnetoelectric coupling}

  The electronic transfer mechanism can be quantitively 
estimated through the change in magnetization of LSMO.
  LSMO is a half-metal with only Mn 3$d$ $e_g$ majority spin levels around 
the Fermi level. 
  As the screening carriers are fully spin-polarized, an electronic transfer between LMSO 
and the BTO surface results in a systematic change of the magnetization near the interface. 
  We calculate the change in magnetization from the pristine to the O-vac and O-ads 
systems and to the 2OH and 2H systems, $\Delta M$, as $\pm$1.7 $\mu_B$
and $\pm$1.5 $\mu_B$ in the supercell, equivalent to $\pm$0.42 $e/S$
and $\pm$0.37 $e/S$ respectively
(the remaining 0.1 electrons/holes stay in BTO, see Appendix).
  This extra electron density (which
corresponds to the electric displacement, $D$, because of the 
half-metallic nature of LSMO) 
resides in the interface region, decaying into the electrode
with an associated Thomas-Fermi screening length (see Fig. 3).
  This situation is similar to the carrier-mediated magnetoelectricity already 
predicted at SrTiO$_3$/SrRuO$_3$ interfaces~\cite{Rondinelli2007}
and in LSMO/BTO superlattices~\cite{Burton2009}.
  In agreement with Ref.~\onlinecite{Burton2009} a competing interface antiferromagnetic type phase 
(called A$_1$ in Table 1 of Ref.~\onlinecite{Burton2009}) was found for the outwards BTO polarization.
   A similar magnetoelectric effect has recently been experimentally 
realized~\cite{Vaz2010,Molegraaf2009}.


\subsection{Tunneling Electroresistance}

  We now discuss how the electrochemical switching process may lead to the
giant TER observed in the LSMO/BTO system~\cite{Garcia2009}.
In the simplest semiclassical approximation, 
  TER has an exponential dependence on the tunneling barrier shape~\cite{Gruverman2009}. 
  The interface dipole, and hence band offset ($E_{\mathrm{VO}}=E_{\mathrm{VBM}}-E_F$),
at a metal/ferroelectric interface depends linearly on the electric displacement field, $D$,
in a way that can be expressed with an effective screening length~\cite{Junquera2003a,Zhuravlev2005,Kohlstedt2005}, $\lambda_{eff}$.
  For LSMO-BTO we calculate $\lambda_{eff}=0.11$ \AA.
%
%
  Using the calculated values of the band offset (Fig. 4 inset) and the 
  experimental band gap of BTO, we obtain the change in 
barrier height upon complete polarization reversal (for $D=\pm P_0$ the 
potential in BTO is flat, i.e. the tunneling barrier shape is rectangular), 
$\Delta \varphi$, and the average barrier height,
$\bar\varphi=(\varphi_{\mathrm{out}}+\varphi_{\mathrm{in}})/2$.
 These values then yield an estimate of the TER using the exponential
dependence~\cite{Gruverman2009} on the barrier thickness, $d$, for large TER,
\begin{equation}
\mathrm{TER} \approx \mathrm{exp}\left[\frac{\sqrt{2m}}{\hbar}\frac{\Delta\varphi}{\sqrt{\bar{\varphi}}}d\right].
\end{equation}
  Fig. 4 compares this estimate with the experimental data~\cite{Garcia2009}
showing that this simple model captures remarkably well the essential 
physics of TER in this system. 
  We note a recent study reported comparable shifts in $E_{\mathrm{VO}}$ (measured using 
  photoelectron spectroscopy) on a similar ferroelectric/LSMO system upon polarization 
  reversal~\cite{Wu2011}.  
  The origin of electroresistance effects in oxide nanotubes has also recently
been suggested as redox reactions~\cite{Nonnenmann2010}. 
  However the redox arguments there are fundamentally different - it is proposed that 
the electrons yielded by oxygen vacancies are directly available for conduction.

\section{Conclusions}

  In conclusion we have studied an electrochemical mechanism
for ferroelectric switching in thin films and proposed it as the origin of
switchable ferroelectricity, TER and magnetoelectricity 
in a prototypical system. 
  This work opens several avenues for future research. 
From the experimental point of view, it would be interesting to investigate 
the composition of a ferroelectric surface before and after switching
(e.g. via the AFM tip), to verify whether reduced or oxidized gas-phase 
species are present (as suggested by our results). 
Also, this point could be indirectly checked by performing the AFM-mediated switching 
experiments in a controlled atmosphere, in analogy to the experiments
of Bi et al. [19] on LAO/STO. 
From the theoretical point of view, a natural next step would be to 
perform a more detailed thermodynamic analysis of the stability of a 
ferroelectric surface (either pristine or decorated with adsorbates).
This would involve exploring different coverages, possible inhomogeneous
polarization states, and the effect of temperature and other external
perturbations. 
We hope that our results will stimulate further 
investigations along these (and possibly other) directions.

\begin{acknowledgments}
  We acknowledge G Catalan, J \'{I}\~{n}iguez, M Bibes, V Garcia, 
N Mathur, X Moya, J Junquera, C Ocal and S Streiffer for valuable discussions,
the support of EPSRC, NANOSELECT and MCINN FIS2009-12721-C04-01
and computing resources of CamGRID at Cambridge, the 
Spanish Supercomputer Network and HPC Europa.
PBL acknowledges DOE support under FWP 70069.
\end{acknowledgments}

\begin{figure*}
\centering
\begin{minipage}{.32\textwidth}
\includegraphics[width=\textwidth]{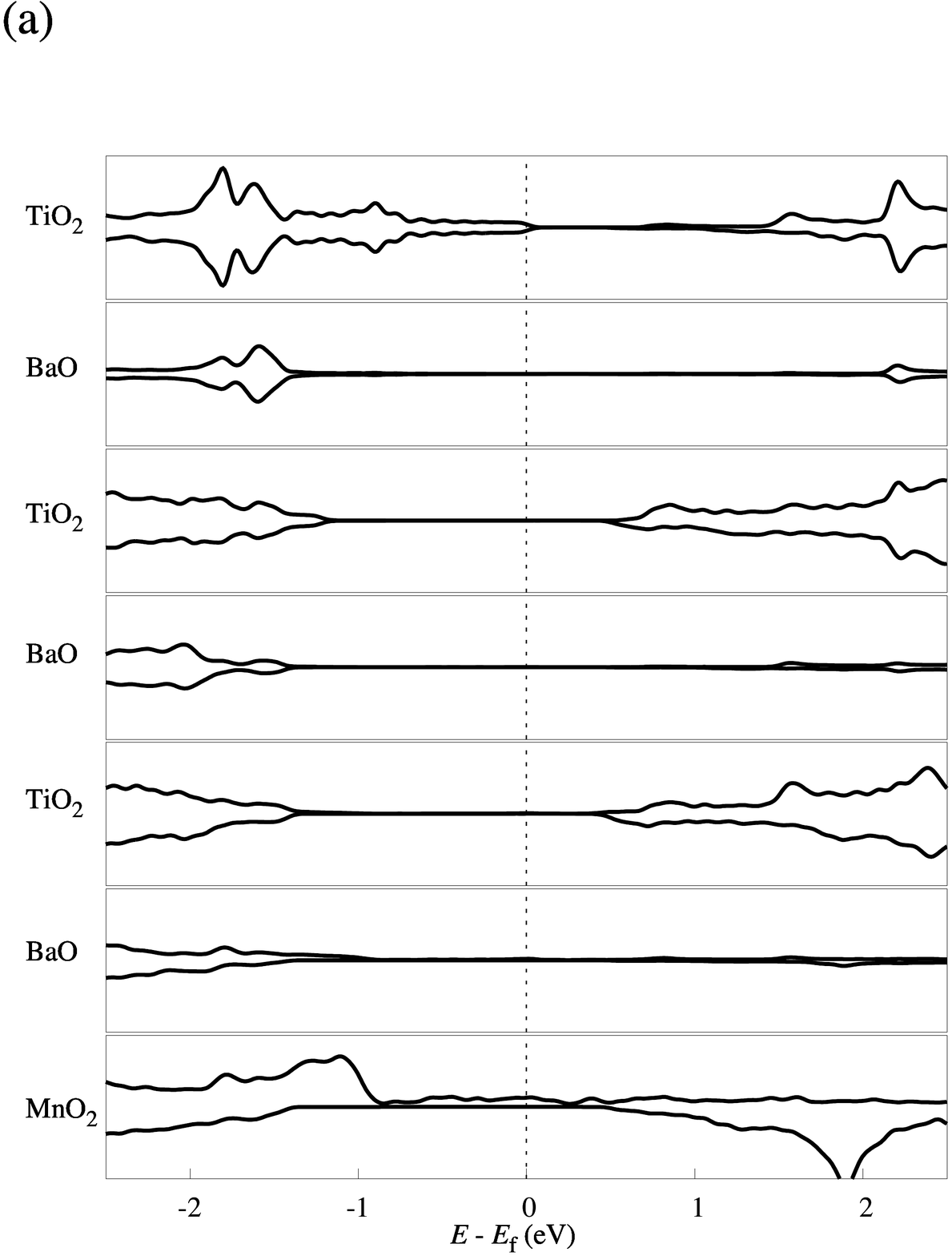}
\end{minipage}
\begin{minipage}{.32\textwidth}
\includegraphics[width=\textwidth]{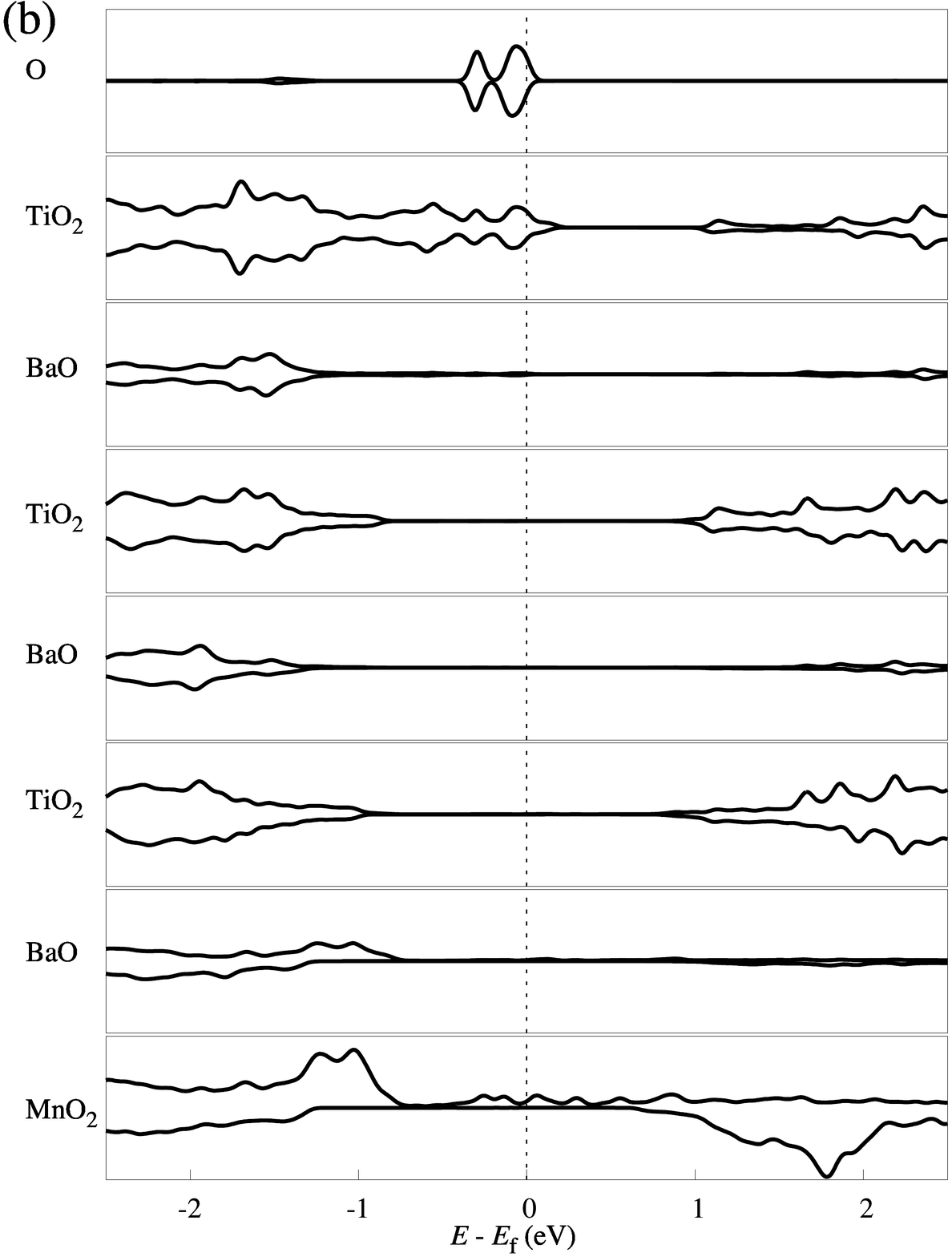}
\end{minipage}
\begin{minipage}{.32\textwidth}
\includegraphics[width=\textwidth]{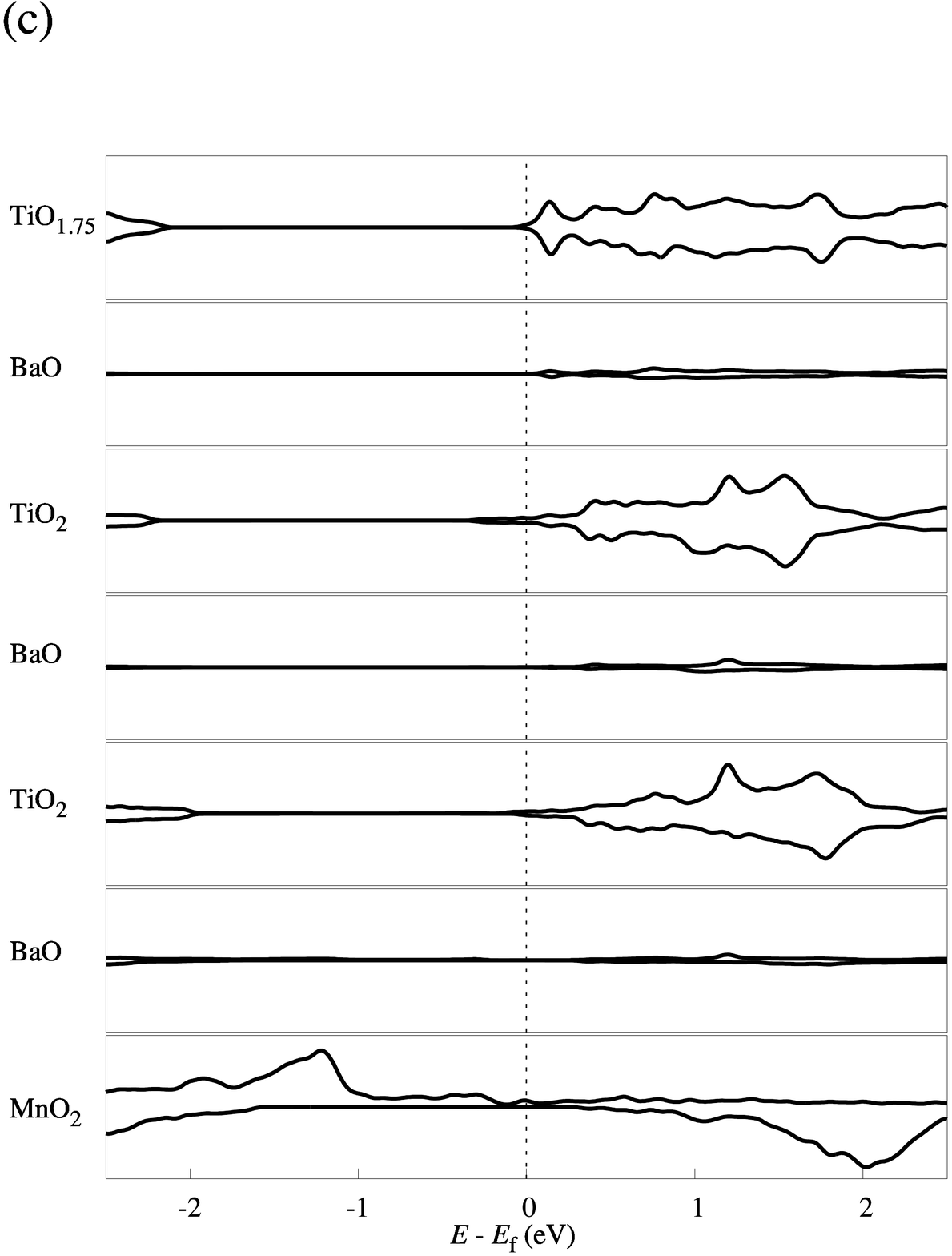}
\end{minipage}
\begin{minipage}{.32\textwidth}
\includegraphics[width=\textwidth]{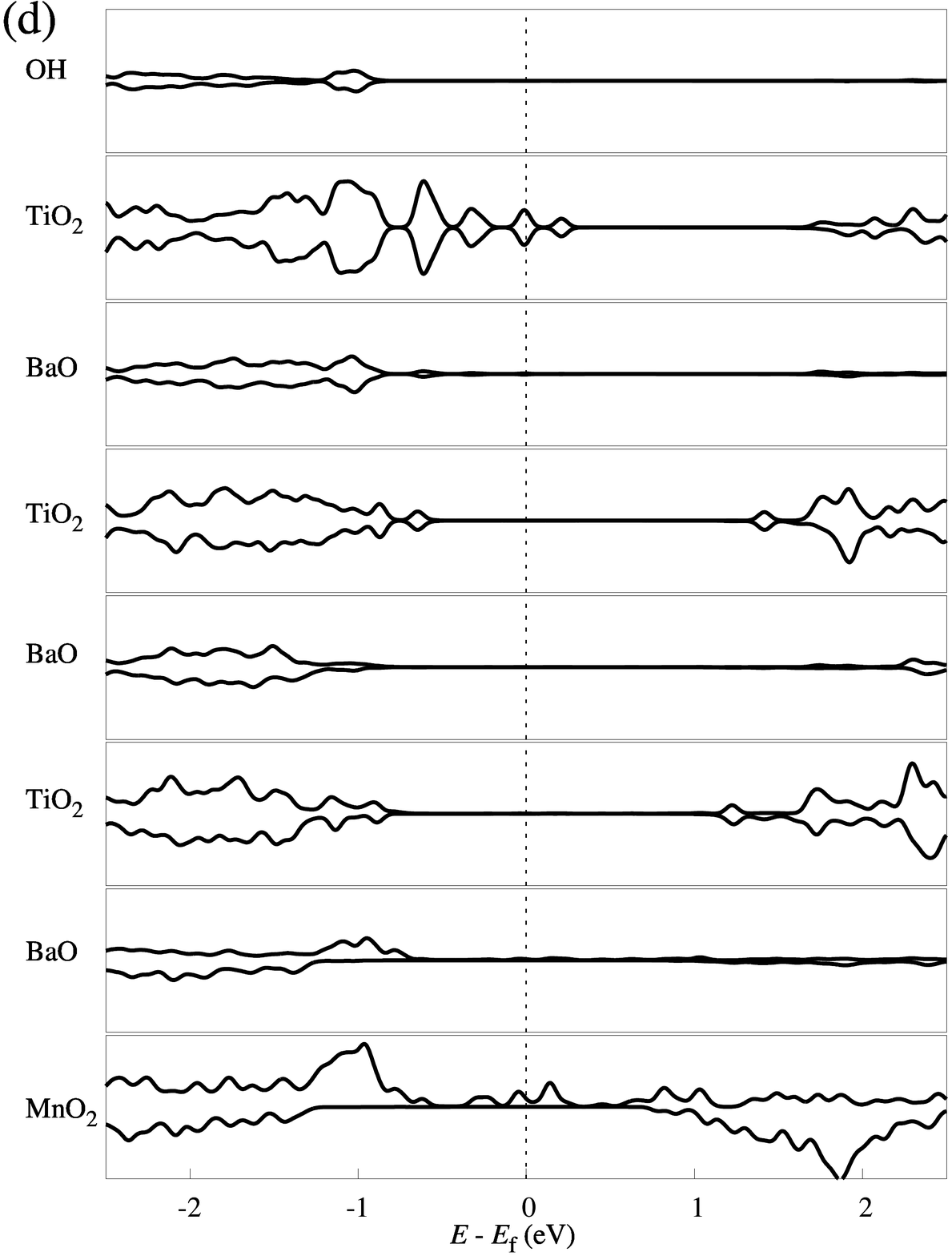}
\end{minipage}
\begin{minipage}{.32\textwidth}
\includegraphics[width=\textwidth]{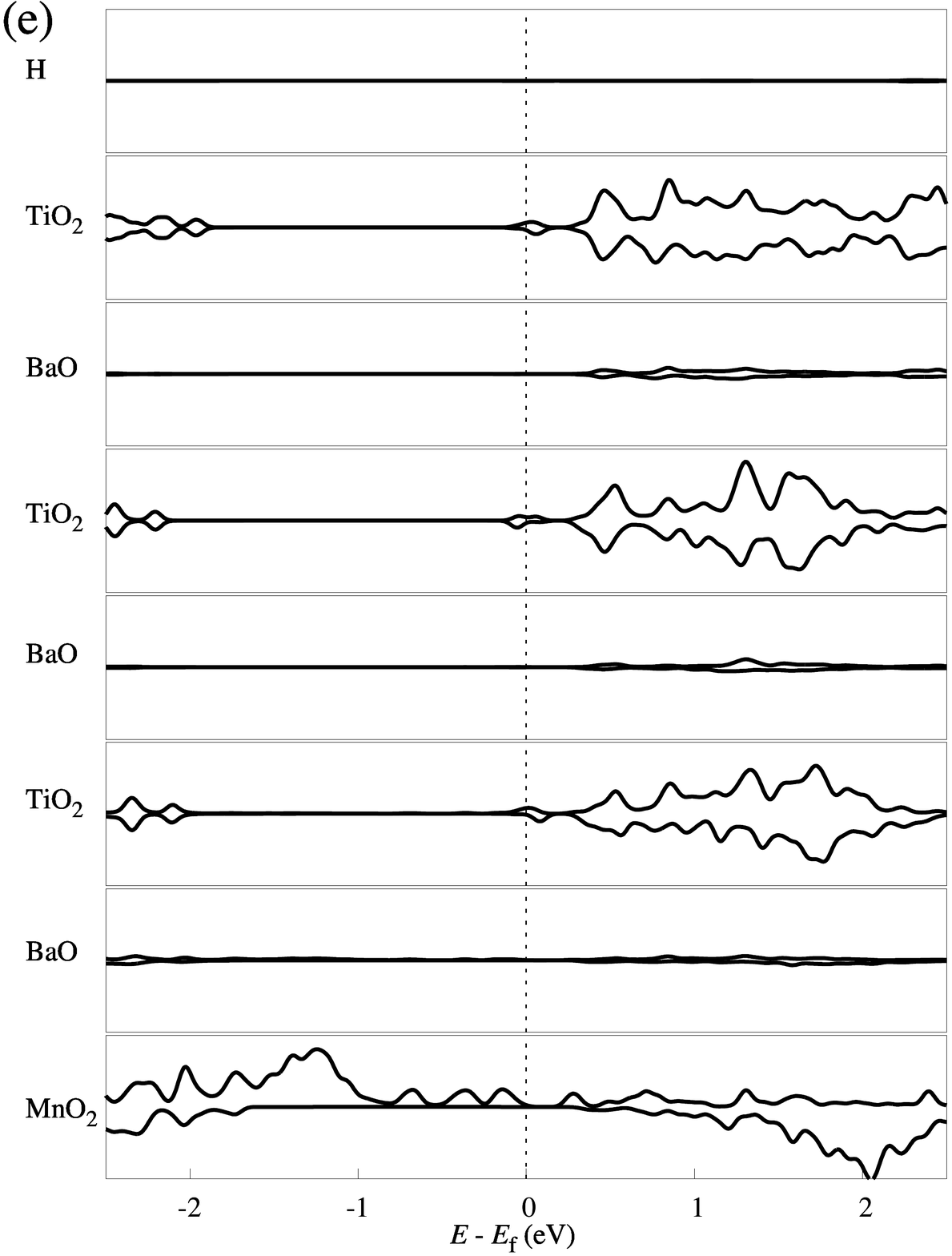}
\end{minipage}
\begin{minipage}{.32\textwidth}
\includegraphics[width=\textwidth]{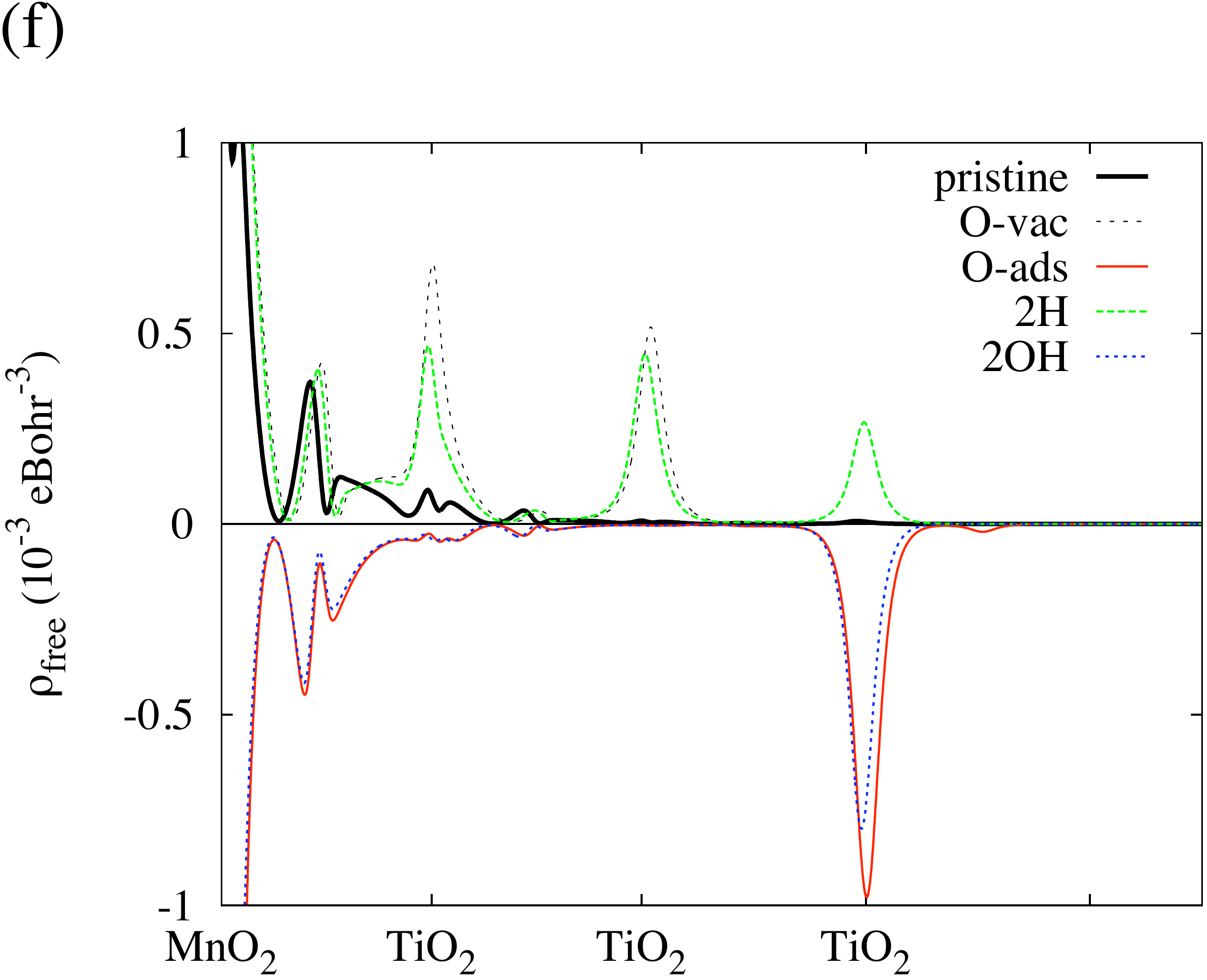}
\end{minipage}
\caption{(Color online) Spin-resolved layer-by-layer density of states centered around $E_\mathrm{f}$ for the LSMO/BTO systems. 
Positive DOS represents majority spin, negative DOS minority spin. Only the BTO layers (and 1 LSMO layer)
are shown for clarity.
Panels (a)-(e) correspond to pristine, O-ads, O-vac, 2OH and 2H systems respectively.
Panel (f) shows a profile of $\rho_{free}$ through the various LSMO/BTO systems.}
\end{figure*}

\appendix*
\section{}

Here we provide details of the electronic structure of the various LSMO/BTO systems.
Figure 5 (a)-(e) shows the spin-resolved layer-by-layer density of states for the pristine, O-ads, O-vac, 2OH and 2H systems.
As discussed in the main text, the electric displacement and polarization within BTO, $D$ and $P$, and hence the valence band offset, $E_{VO}$, depend only on
$Q/S$, the surface defect charge density, and not the surface chemistry. 
However, as discussed in Ref.~\onlinecite{Stengel2011a}, this is not strictly the case once $E_{VO}$ becomes negative or reaches 
the band gap of BTO. 
At this point electrons or holes ``spill out" in to the BTO layer. 
This ``charge spill out" regime is favored by DFT, which often underestimates the band gap, and therefore can be an artifact of the calculation. 

\begin{figure*}[t]
\centering
\begin{minipage}{.2\textwidth}
\includegraphics[width=\textwidth]{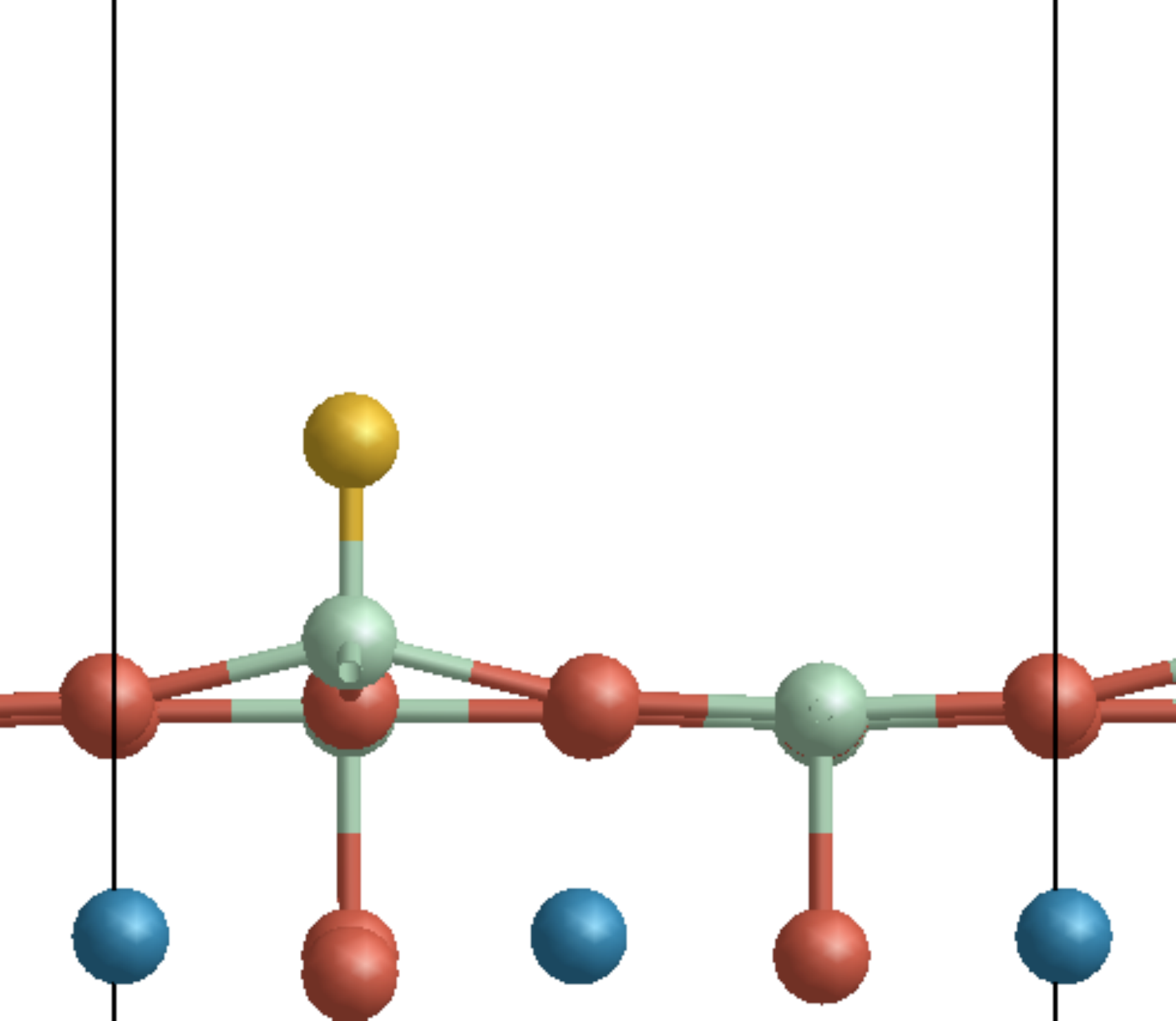}
\end{minipage}
\begin{minipage}{.2\textwidth}
\includegraphics[width=\textwidth]{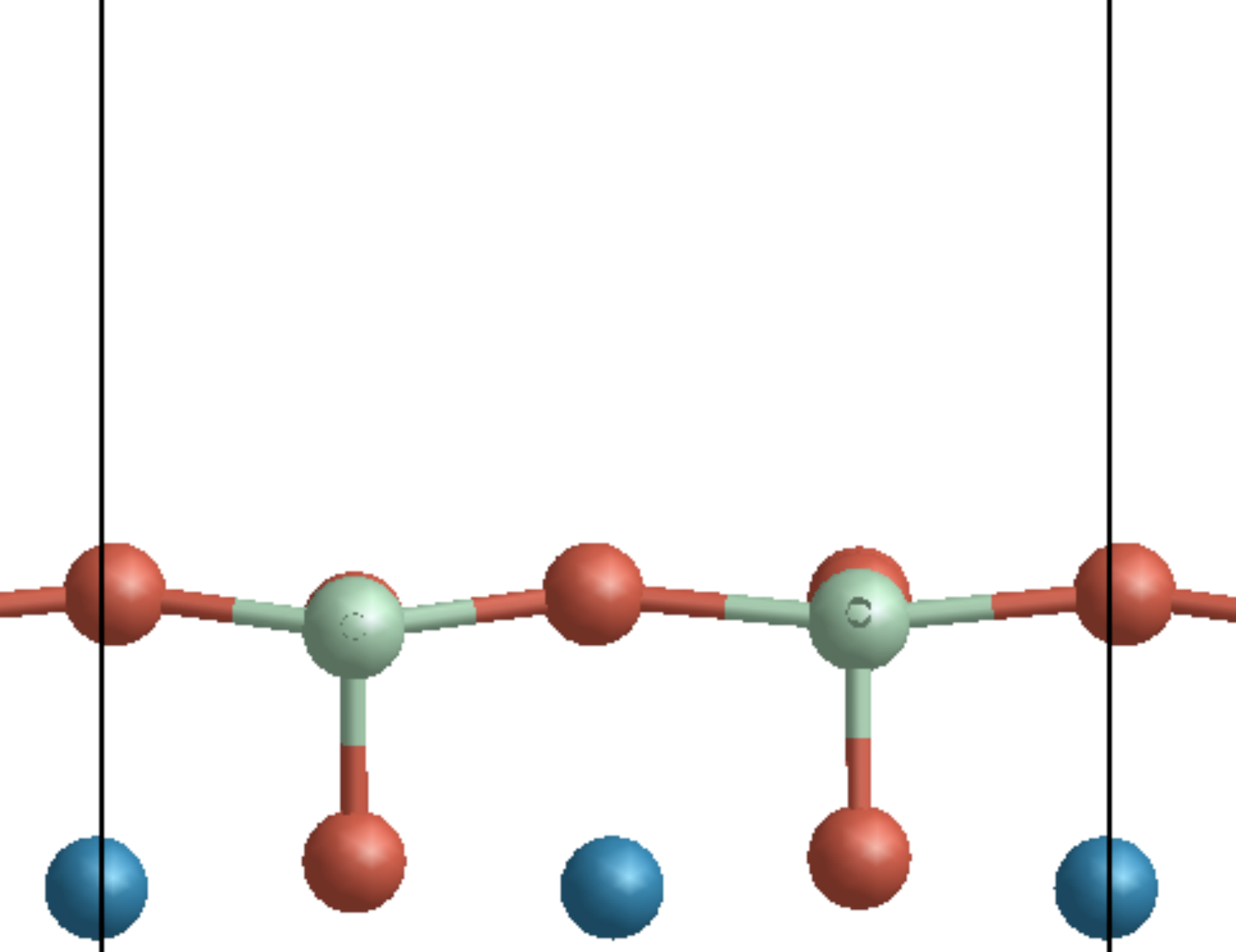}
\end{minipage}
\begin{minipage}{.2\textwidth}
\includegraphics[width=\textwidth]{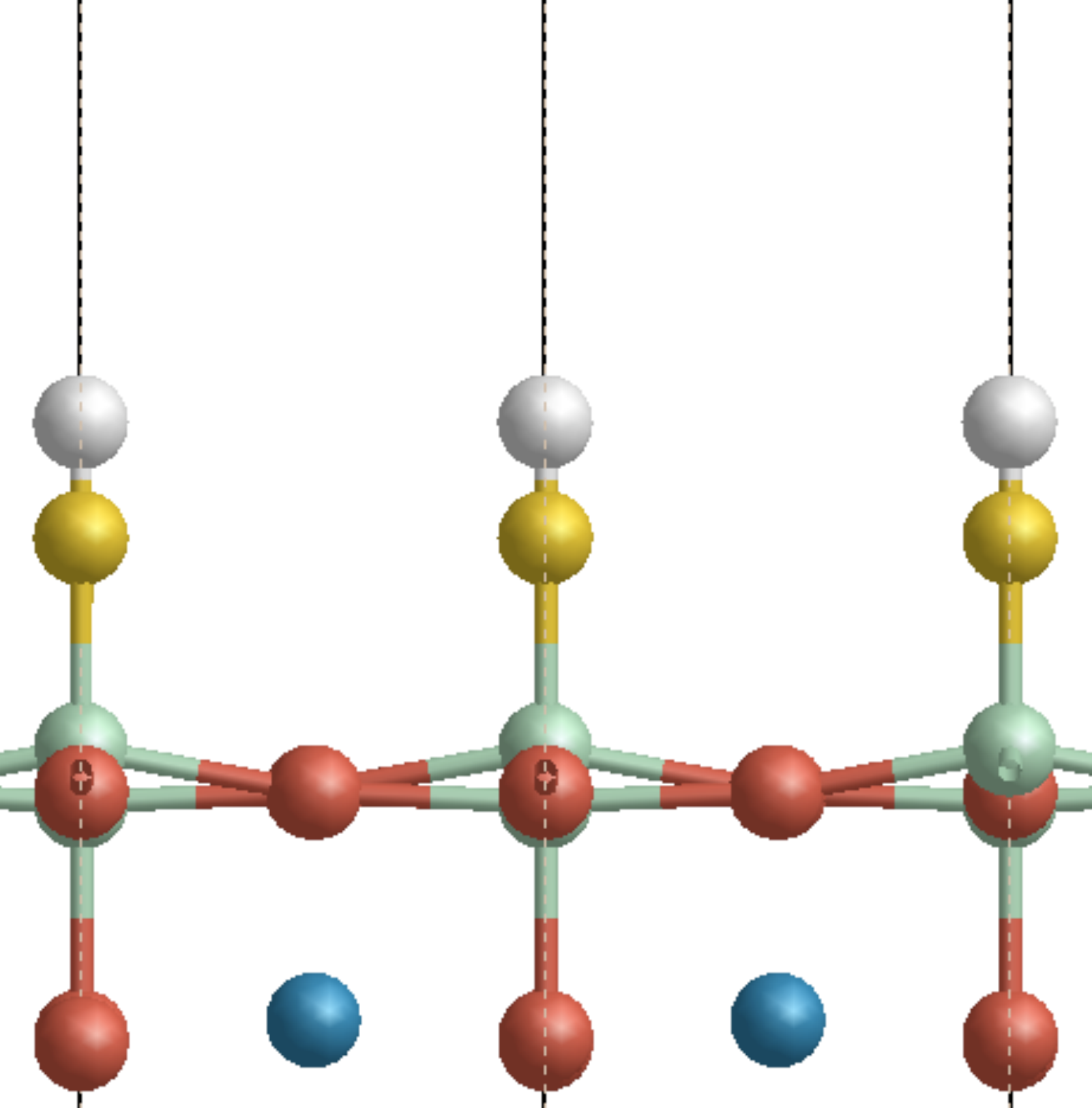}
\end{minipage}
\begin{minipage}{.2\textwidth}
\includegraphics[width=\textwidth]{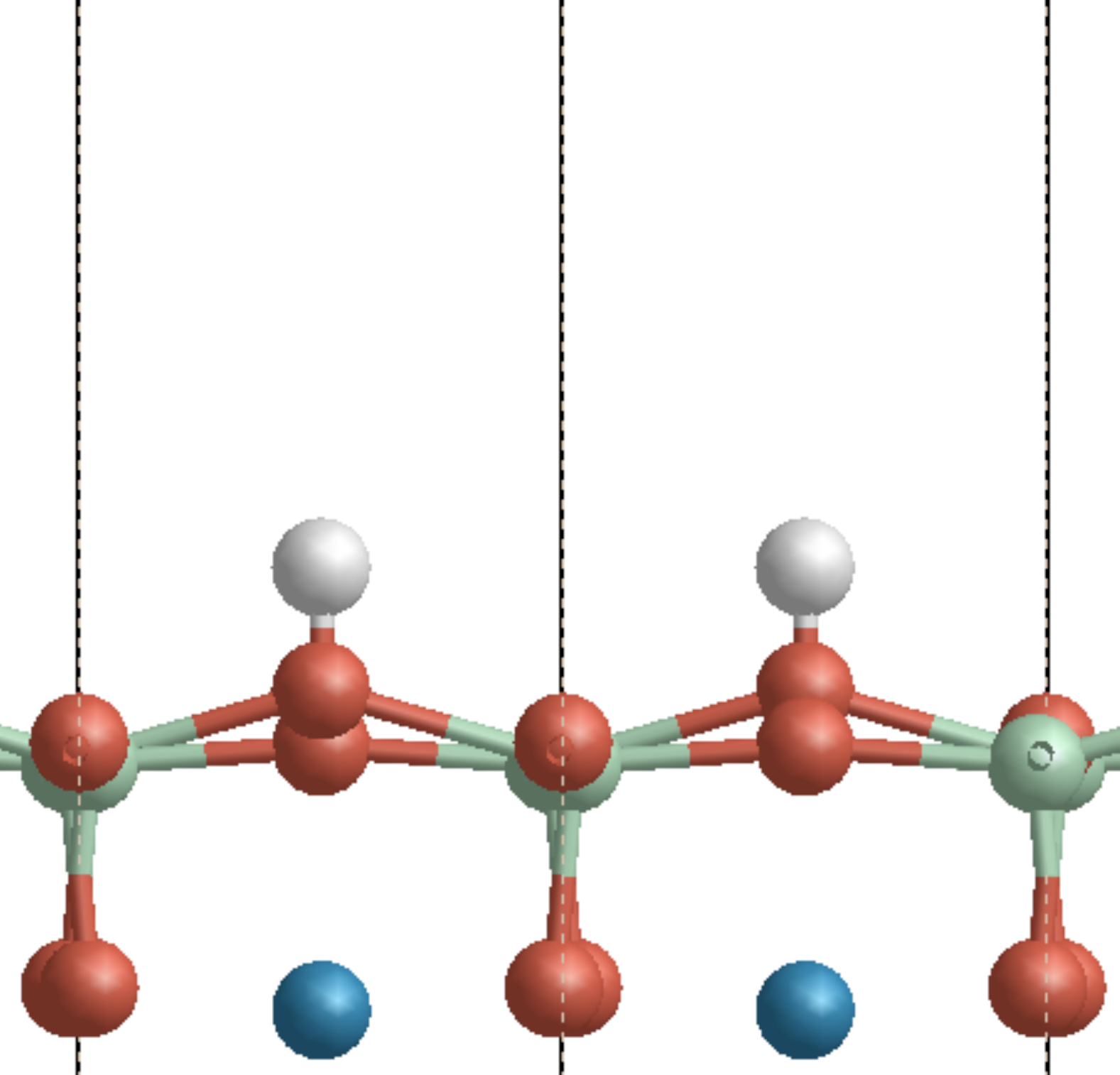}
\end{minipage}
\begin{minipage}{.2\textwidth}
\includegraphics[width=\textwidth]{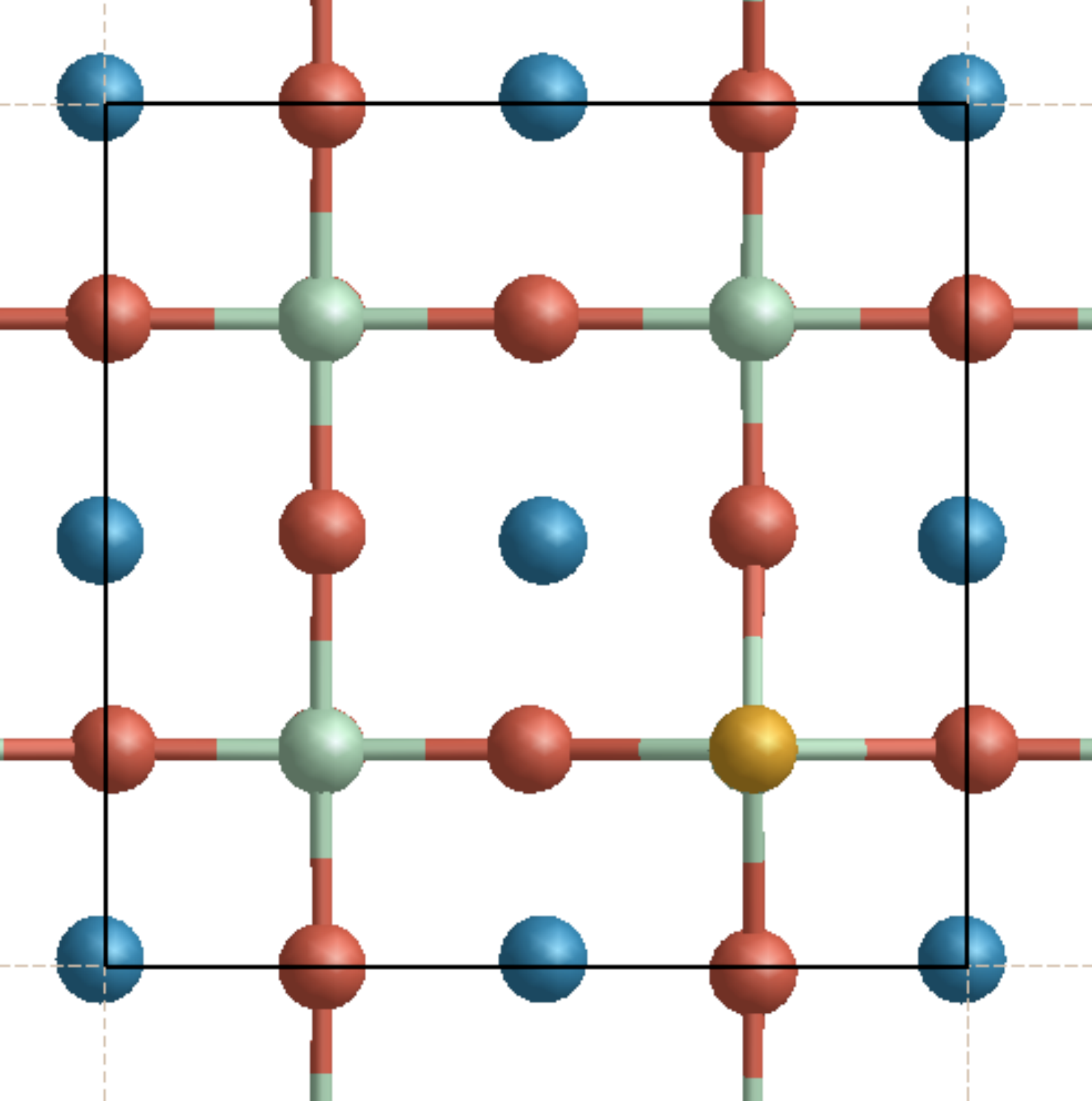}
\end{minipage}
\begin{minipage}{.2\textwidth}
\includegraphics[width=\textwidth]{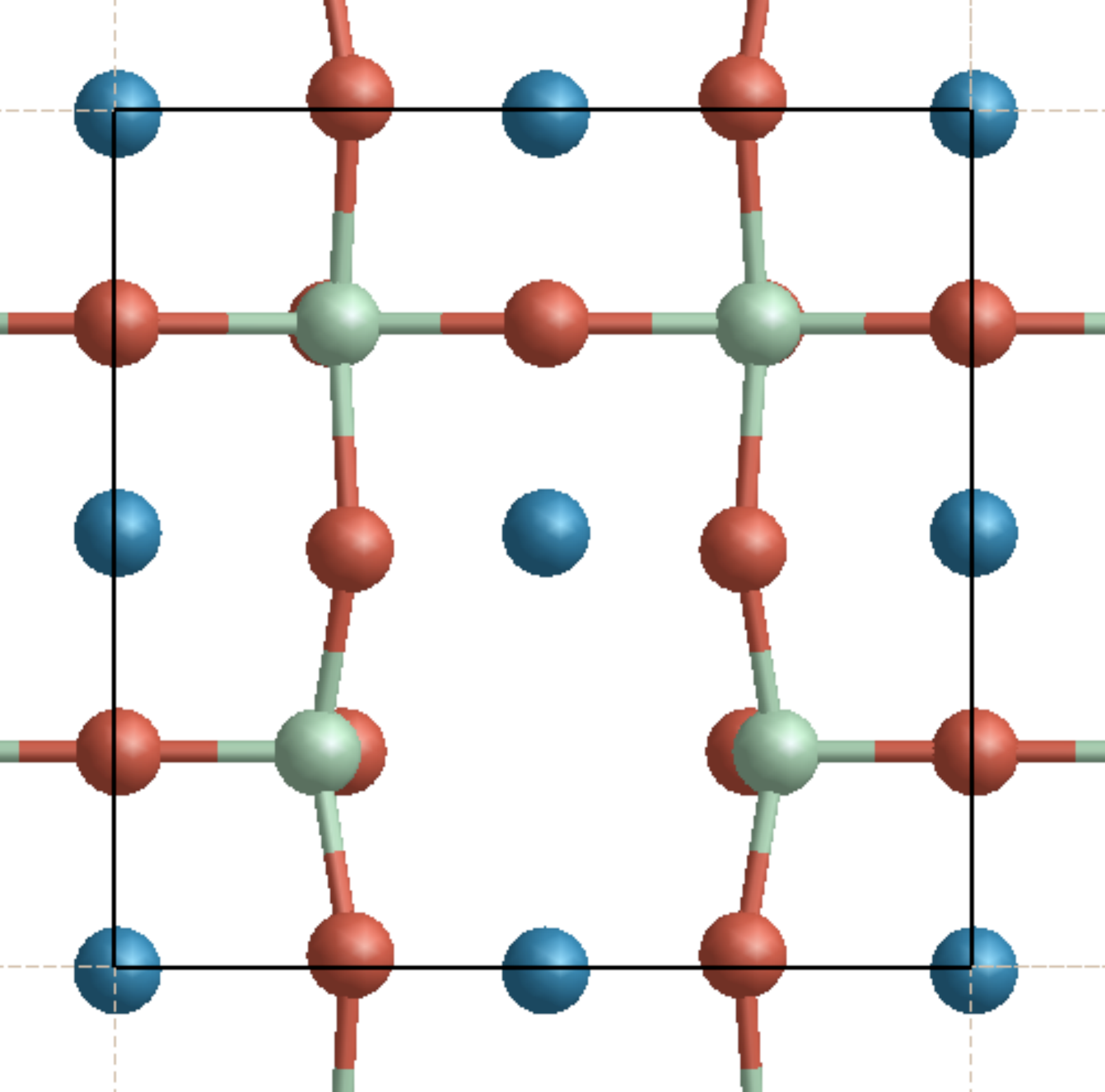}
\end{minipage}
\begin{minipage}{.2\textwidth}
\includegraphics[width=\textwidth]{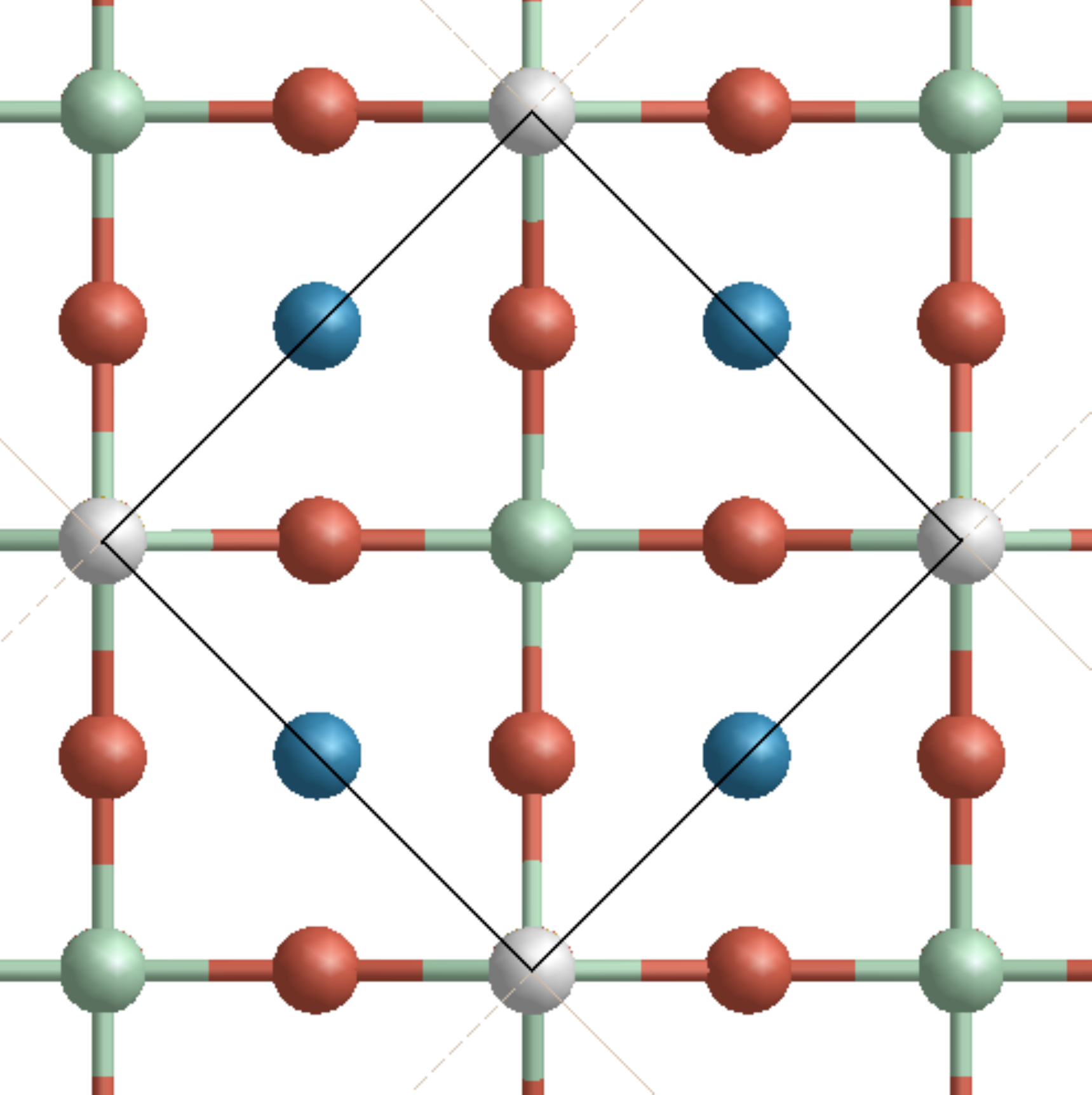}
\end{minipage}
\begin{minipage}{.2\textwidth}
\includegraphics[width=\textwidth]{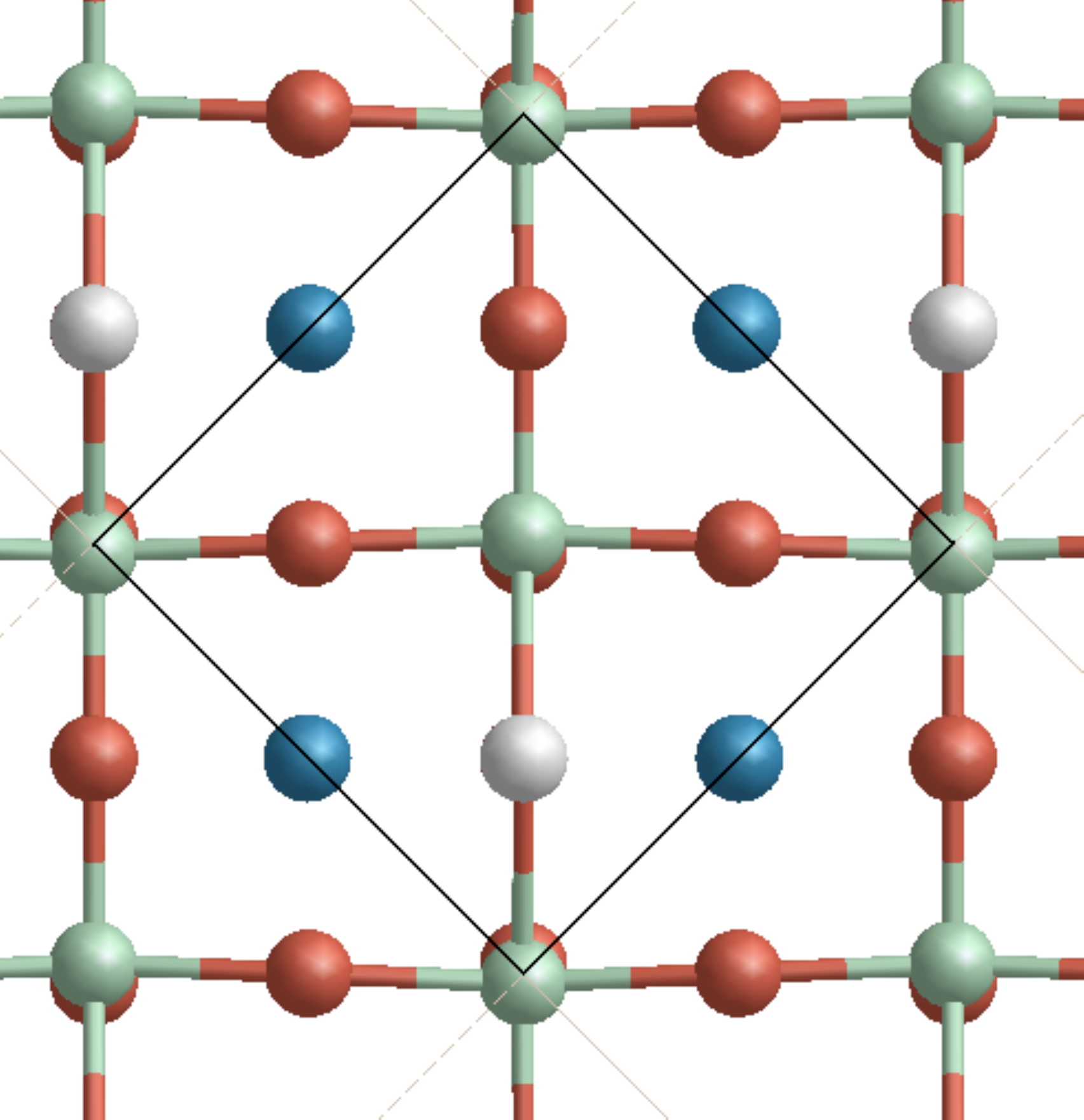}
\end{minipage}
\caption{Relaxed surface LSMO/BTO structure. 
Top: view along the [100] direction.
Bottom: birds-eye view along the [001] direction. 
 Only the top BTO layer is shown for clarity. Sr (blue), Ti (cyan), O (red), O-ads (orange).
 Panels correspond to O-ads, O-vac, 2OH and 2H systems respectively (left to right).}
\end{figure*}

In our case, due to the presence of a free surface, there is
a further issue that was not explicitly considered in Ref. 35,
i.e. the effect of surface states. 
In many cases, these fall within the bulk band gap of the ferroelectric film, and might
cross the Fermi level of the metal, thus causing a significant
spill-out of charge even when the bulk electronic bands are not
directly affected. 
Note that surface states in ferroelectrics
typically have a marked localized orbital character (either the
atomic orbital of an adsorbate, or the 3$d$ orbitals of the 
transition metal cation). 
Therefore, it is reasonable to suspect 
that DFT might introduce systematic errors in their ionization
energies (similarly to the energy location of the bulk band edges 
discussed in Ref. 35), and the metallization of a surface state
should be regarded with analogous caution 
(for a detailed 
discussion of charge transfers at surfaces see, e.g., Ref.~\onlinecite{Stengel2011b}).

Refs.~\onlinecite{Stengel2011a} and~\onlinecite{Stengel2011b} prescribe an analysis of the hole- and electron-like charge spill out. 
Following this prescription, we determine the free electron density profile, $\rho_{free}$, within BTO in Fig. 5(f) (using Eq. 25 and 26 of Ref.~\onlinecite{Stengel2011a} and 
Eq. 19 of Ref.~\onlinecite{Stengel2011b} for hole spill out).
Out of all the five systems, the pristine one is unaffected, the negatively polarized
(O-vac and 2H) systems are affected by electron spill out into
the conduction band, and the two positively polarized (O-ads and 2OH)
ones are affected by hole spill out into surface states (see Fig. 5).
In both O-vac and 2H $\rho_{free}$ amounts to approximately 0.03 electrons per unit cell of BTO
, which is a fairly mild effect (compare with approximately 0.15 electrons per unit cell in the KNO/SRO system of Ref.~\onlinecite{Stengel2011a}).
In the case of O-ads and 2OH, the surface O(2$p$) states accommodate a total of approximately 0.1 holes per surface perovskite unit (Fig. 5(f)).
Of course, estimating to what degree this charge spill is problematic, depends not only on the magnitude but also the purpose of the calculation. 
The charge spill out induces an error in two quantities that are discussed in
this work: the total injected charge into the LSMO electrode,
and the band alignment. 
Considering the total injected charge,
the impact of this error is trivial to estimate. 
In fact, 0.03 electrons per cell times $N$, number of BTO cells, corresponds  exactly
to the difference between the actual induced spin in LSMO and the ``ideal" limit of 2 Bohr magnetons per cell. 
This observation can be directly used to estimate the error in the 
calculated band offset. 
In fact, we can assume in a first approximation
that the band offset is linear in the electric displacement of 
the BTO cell adjacent to the interface, $D_{inter}$. 
Using the above numbers for the O-vac system,
\begin{equation}
D_{inter} = 2e/S - 0.03e(N/S) = M/S = 1.7e/S,
\end{equation}
where $M$ is the induced spin, and $S$ is the supercell surface area (or reciprocal of the defect density).
This provides an accurate estimate of the actual electric 
displacement ``felt" by LSMO. 
Using this information, therefore,
we can make a very accurate estimation of the linear band offset
dependence with $D$. 
We used this analysis to make the plot shown
in Fig. 4 inset of the main text.
Therefore, whilst we understand the limitations of DFT, 
in this case they do not affect significantly our conclusions.


Figure 6 displays the relaxed atomic structures of the BTO surface with O-ads, O-vac, 2OH and 2H.



\begin{thebibliography}{38}
\expandafter\ifx\csname natexlab\endcsname\relax\def\natexlab#1{#1}\fi
\expandafter\ifx\csname bibnamefont\endcsname\relax
  \def\bibnamefont#1{#1}\fi
\expandafter\ifx\csname bibfnamefont\endcsname\relax
  \def\bibfnamefont#1{#1}\fi
\expandafter\ifx\csname citenamefont\endcsname\relax
  \def\citenamefont#1{#1}\fi
\expandafter\ifx\csname url\endcsname\relax
  \def\url#1{\texttt{#1}}\fi
\expandafter\ifx\csname urlprefix\endcsname\relax\def\urlprefix{URL }\fi
\providecommand{\bibinfo}[2]{#2}
\providecommand{\eprint}[2][]{\url{#2}}

\bibitem[{\citenamefont{Garcia et~al.}(2009)\citenamefont{Garcia, Fusil,
  Bouzehouane, Enouz-Vedrenne, Mathur, Barth{\'e}l{\'e}my, and
  Bibes}}]{Garcia2009}
\bibinfo{author}{\bibfnamefont{V.}~\bibnamefont{Garcia}},
  \bibinfo{author}{\bibfnamefont{S.}~\bibnamefont{Fusil}},
  \bibinfo{author}{\bibfnamefont{K.}~\bibnamefont{Bouzehouane}},
  \bibinfo{author}{\bibfnamefont{S.}~\bibnamefont{Enouz-Vedrenne}},
  \bibinfo{author}{\bibfnamefont{N.}~\bibnamefont{Mathur}},
  \bibinfo{author}{\bibfnamefont{A.}~\bibnamefont{Barth{\'e}l{\'e}my}},
  \bibnamefont{and} \bibinfo{author}{\bibfnamefont{M.}~\bibnamefont{Bibes}},
  \bibinfo{journal}{Nature} \textbf{\bibinfo{volume}{460}}, \bibinfo{pages}{81}
  (\bibinfo{year}{2009}).

\bibitem[{\citenamefont{Zubko and Triscone}(2009)}]{Zubko2009}
\bibinfo{author}{\bibfnamefont{P.}~\bibnamefont{Zubko}} \bibnamefont{and}
  \bibinfo{author}{\bibfnamefont{J.}~\bibnamefont{Triscone}},
  \bibinfo{journal}{Nature} \textbf{\bibinfo{volume}{460}}, \bibinfo{pages}{45}
  (\bibinfo{year}{2009}).

\bibitem[{\citenamefont{Segal}(2009)}]{Segal2009}
\bibinfo{author}{\bibfnamefont{M.}~\bibnamefont{Segal}},
  \bibinfo{journal}{Nature Nanotechnology}  (\bibinfo{year}{2009}).

\bibitem[{\citenamefont{Fridkin}(1980)}]{Fridkin1980}
\bibinfo{author}{\bibfnamefont{V.}~\bibnamefont{Fridkin}},
  \emph{\bibinfo{title}{Ferroelectric semiconductors}}
  (\bibinfo{publisher}{Consultants Bureau}, \bibinfo{year}{1980}), ISBN
  \bibinfo{isbn}{0306109573}.

\bibitem[{\citenamefont{Wang et~al.}(2009)\citenamefont{Wang, Fong, Jiang,
  Highland, Fuoss, Thompson, Kolpak, Eastman, Streiffer, Rappe
  et~al.}}]{Wang2009}
\bibinfo{author}{\bibfnamefont{R.}~\bibnamefont{Wang}},
  \bibinfo{author}{\bibfnamefont{D.}~\bibnamefont{Fong}},
  \bibinfo{author}{\bibfnamefont{F.}~\bibnamefont{Jiang}},
  \bibinfo{author}{\bibfnamefont{M.}~\bibnamefont{Highland}},
  \bibinfo{author}{\bibfnamefont{P.}~\bibnamefont{Fuoss}},
  \bibinfo{author}{\bibfnamefont{C.}~\bibnamefont{Thompson}},
  \bibinfo{author}{\bibfnamefont{A.}~\bibnamefont{Kolpak}},
  \bibinfo{author}{\bibfnamefont{J.}~\bibnamefont{Eastman}},
  \bibinfo{author}{\bibfnamefont{S.}~\bibnamefont{Streiffer}},
  \bibinfo{author}{\bibfnamefont{A.}~\bibnamefont{Rappe}},
  \bibnamefont{et~al.}, \bibinfo{journal}{Phys. Rev. Lett.}
  \textbf{\bibinfo{volume}{102}}, \bibinfo{pages}{47601}
  (\bibinfo{year}{2009}).

\bibitem[{\citenamefont{Fong et~al.}(2006)\citenamefont{Fong, Kolpak, Eastman,
  Streiffer, Fuoss, Stephenson, Thompson, Kim, Choi, Eom et~al.}}]{Fong2006}
\bibinfo{author}{\bibfnamefont{D.}~\bibnamefont{Fong}},
  \bibinfo{author}{\bibfnamefont{A.}~\bibnamefont{Kolpak}},
  \bibinfo{author}{\bibfnamefont{J.}~\bibnamefont{Eastman}},
  \bibinfo{author}{\bibfnamefont{S.}~\bibnamefont{Streiffer}},
  \bibinfo{author}{\bibfnamefont{P.}~\bibnamefont{Fuoss}},
  \bibinfo{author}{\bibfnamefont{G.}~\bibnamefont{Stephenson}},
  \bibinfo{author}{\bibfnamefont{C.}~\bibnamefont{Thompson}},
  \bibinfo{author}{\bibfnamefont{D.}~\bibnamefont{Kim}},
  \bibinfo{author}{\bibfnamefont{K.}~\bibnamefont{Choi}},
  \bibinfo{author}{\bibfnamefont{C.}~\bibnamefont{Eom}}, \bibnamefont{et~al.},
  \bibinfo{journal}{Phys. Rev. Lett.} \textbf{\bibinfo{volume}{96}},
  \bibinfo{pages}{127601} (\bibinfo{year}{2006}).

\bibitem[{\citenamefont{Spanier et~al.}(2006)\citenamefont{Spanier, Kolpak,
  Urban, Grinberg, Ouyang, Yun, Rappe, and Park}}]{Spanier2006}
\bibinfo{author}{\bibfnamefont{J.}~\bibnamefont{Spanier}},
  \bibinfo{author}{\bibfnamefont{A.}~\bibnamefont{Kolpak}},
  \bibinfo{author}{\bibfnamefont{J.}~\bibnamefont{Urban}},
  \bibinfo{author}{\bibfnamefont{I.}~\bibnamefont{Grinberg}},
  \bibinfo{author}{\bibfnamefont{L.}~\bibnamefont{Ouyang}},
  \bibinfo{author}{\bibfnamefont{W.}~\bibnamefont{Yun}},
  \bibinfo{author}{\bibfnamefont{A.}~\bibnamefont{Rappe}}, \bibnamefont{and}
  \bibinfo{author}{\bibfnamefont{H.}~\bibnamefont{Park}},
  \bibinfo{journal}{Nano Lett.} \textbf{\bibinfo{volume}{6}},
  \bibinfo{pages}{735} (\bibinfo{year}{2006}).

\bibitem[{\citenamefont{Cen et~al.}(2008)\citenamefont{Cen, Thiel, Hammerl,
  Schneider, Andersen, Hellberg, Mannhart, and Levy}}]{Cen2008}
\bibinfo{author}{\bibfnamefont{C.}~\bibnamefont{Cen}},
  \bibinfo{author}{\bibfnamefont{S.}~\bibnamefont{Thiel}},
  \bibinfo{author}{\bibfnamefont{G.}~\bibnamefont{Hammerl}},
  \bibinfo{author}{\bibfnamefont{C.~W.} \bibnamefont{Schneider}},
  \bibinfo{author}{\bibfnamefont{K.~E.} \bibnamefont{Andersen}},
  \bibinfo{author}{\bibfnamefont{C.~S.} \bibnamefont{Hellberg}},
  \bibinfo{author}{\bibfnamefont{J.}~\bibnamefont{Mannhart}}, \bibnamefont{and}
  \bibinfo{author}{\bibfnamefont{J.}~\bibnamefont{Levy}},
  \bibinfo{journal}{Nat. Mater.} \textbf{\bibinfo{volume}{7}},
  \bibinfo{pages}{298} (\bibinfo{year}{2008}).

\bibitem[{\citenamefont{Bristowe et~al.}(2011)\citenamefont{Bristowe,
  Littlewood, and Artacho}}]{Bristowe2011a}
\bibinfo{author}{\bibfnamefont{N.~C.} \bibnamefont{Bristowe}},
  \bibinfo{author}{\bibfnamefont{P.~B.} \bibnamefont{Littlewood}},
  \bibnamefont{and} \bibinfo{author}{\bibfnamefont{E.}~\bibnamefont{Artacho}},
  \bibinfo{journal}{Phys. Rev. B} \textbf{\bibinfo{volume}{83}},
  \bibinfo{pages}{205405} (\bibinfo{year}{2011}).

\bibitem[{\citenamefont{Xie et~al.}(2010)\citenamefont{Xie, Bell, Yajima,
  Hikita, and Hwang}}]{Xie2010}
\bibinfo{author}{\bibfnamefont{Y.}~\bibnamefont{Xie}},
  \bibinfo{author}{\bibfnamefont{C.}~\bibnamefont{Bell}},
  \bibinfo{author}{\bibfnamefont{T.}~\bibnamefont{Yajima}},
  \bibinfo{author}{\bibfnamefont{Y.}~\bibnamefont{Hikita}}, \bibnamefont{and}
  \bibinfo{author}{\bibfnamefont{H.}~\bibnamefont{Hwang}},
  \bibinfo{journal}{Nano Lett.}  (\bibinfo{year}{2010}).

\bibitem[{\citenamefont{Wu and Cohen}(2006)}]{Wu2006a}
\bibinfo{author}{\bibfnamefont{Z.}~\bibnamefont{Wu}} \bibnamefont{and}
  \bibinfo{author}{\bibfnamefont{R.}~\bibnamefont{Cohen}},
  \bibinfo{journal}{Phys. Rev. B} \textbf{\bibinfo{volume}{73}},
  \bibinfo{pages}{235116} (\bibinfo{year}{2006}).

\bibitem[{\citenamefont{Ordejon et~al.}(1996)\citenamefont{Ordejon, Artacho,
  and Soler}}]{Ordejon1996}
\bibinfo{author}{\bibfnamefont{P.}~\bibnamefont{Ordejon}},
  \bibinfo{author}{\bibfnamefont{E.}~\bibnamefont{Artacho}}, \bibnamefont{and}
  \bibinfo{author}{\bibfnamefont{J.~M.} \bibnamefont{Soler}},
  \bibinfo{journal}{Phys. Rev. B} \textbf{\bibinfo{volume}{53}},
  \bibinfo{pages}{10441} (\bibinfo{year}{1996}).

\bibitem[{\citenamefont{Soler et~al.}(2002)\citenamefont{Soler, Artacho, Gale,
  Garcia, Junquera, Ordejon, and Sanchez-Portal}}]{Soler2002}
\bibinfo{author}{\bibfnamefont{J.}~\bibnamefont{Soler}},
  \bibinfo{author}{\bibfnamefont{E.}~\bibnamefont{Artacho}},
  \bibinfo{author}{\bibfnamefont{J.}~\bibnamefont{Gale}},
  \bibinfo{author}{\bibfnamefont{A.}~\bibnamefont{Garcia}},
  \bibinfo{author}{\bibfnamefont{J.}~\bibnamefont{Junquera}},
  \bibinfo{author}{\bibfnamefont{P.}~\bibnamefont{Ordejon}}, \bibnamefont{and}
  \bibinfo{author}{\bibfnamefont{D.}~\bibnamefont{Sanchez-Portal}},
  \bibinfo{journal}{J. Phys.: Condens. Matter} \textbf{\bibinfo{volume}{14}},
  \bibinfo{pages}{2745} (\bibinfo{year}{2002}).

\bibitem[{\citenamefont{Ferrari et~al.}(2006)\citenamefont{Ferrari, Pruneda,
  and Artacho}}]{Ferrari2006}
\bibinfo{author}{\bibfnamefont{V.}~\bibnamefont{Ferrari}},
  \bibinfo{author}{\bibfnamefont{J.}~\bibnamefont{Pruneda}}, \bibnamefont{and}
  \bibinfo{author}{\bibfnamefont{E.}~\bibnamefont{Artacho}},
  \bibinfo{journal}{Physica Status Solidi (a)} \textbf{\bibinfo{volume}{203}},
  \bibinfo{pages}{1437} (\bibinfo{year}{2006}).

\bibitem[{\citenamefont{Pruneda et~al.}(2007)\citenamefont{Pruneda, Ferrari,
  Rurali, Littlewood, Spaldin, and Artacho}}]{Pruneda2007}
\bibinfo{author}{\bibfnamefont{J.}~\bibnamefont{Pruneda}},
  \bibinfo{author}{\bibfnamefont{V.}~\bibnamefont{Ferrari}},
  \bibinfo{author}{\bibfnamefont{R.}~\bibnamefont{Rurali}},
  \bibinfo{author}{\bibfnamefont{P.}~\bibnamefont{Littlewood}},
  \bibinfo{author}{\bibfnamefont{N.}~\bibnamefont{Spaldin}}, \bibnamefont{and}
  \bibinfo{author}{\bibfnamefont{E.}~\bibnamefont{Artacho}},
  \bibinfo{journal}{Phys. Rev. Lett.} \textbf{\bibinfo{volume}{99}},
  \bibinfo{pages}{226101} (\bibinfo{year}{2007}).

\bibitem[{\citenamefont{Perdew et~al.}(1996)\citenamefont{Perdew, Burke, and
  Ernzerhof}}]{Perdew1996}
\bibinfo{author}{\bibfnamefont{J.}~\bibnamefont{Perdew}},
  \bibinfo{author}{\bibfnamefont{K.}~\bibnamefont{Burke}}, \bibnamefont{and}
  \bibinfo{author}{\bibfnamefont{M.}~\bibnamefont{Ernzerhof}},
  \bibinfo{journal}{Phys. Rev. Lett.} \textbf{\bibinfo{volume}{77}},
  \bibinfo{pages}{3865} (\bibinfo{year}{1996}).

\bibitem[{\citenamefont{Padilla and Vanderbilt}(1997)}]{Padilla1997}
\bibinfo{author}{\bibfnamefont{J.}~\bibnamefont{Padilla}} \bibnamefont{and}
  \bibinfo{author}{\bibfnamefont{D.}~\bibnamefont{Vanderbilt}},
  \bibinfo{journal}{Phys. Rev. B} \textbf{\bibinfo{volume}{56}},
  \bibinfo{pages}{1625} (\bibinfo{year}{1997}).

\bibitem[{\citenamefont{Heifets et~al.}(1997)\citenamefont{Heifets, Dorfman,
  Fuks, and Kotomin}}]{Heifets1997}
\bibinfo{author}{\bibfnamefont{E.}~\bibnamefont{Heifets}},
  \bibinfo{author}{\bibfnamefont{S.}~\bibnamefont{Dorfman}},
  \bibinfo{author}{\bibfnamefont{D.}~\bibnamefont{Fuks}}, \bibnamefont{and}
  \bibinfo{author}{\bibfnamefont{E.}~\bibnamefont{Kotomin}},
  \bibinfo{journal}{Thin Solid Films} \textbf{\bibinfo{volume}{296}},
  \bibinfo{pages}{76} (\bibinfo{year}{1997}).

\bibitem[{\citenamefont{Bi et~al.}(2010)\citenamefont{Bi, Bogorin, Cen, Bark,
  Park, Eom, and Levy}}]{Bi2010}
\bibinfo{author}{\bibfnamefont{F.}~\bibnamefont{Bi}},
  \bibinfo{author}{\bibfnamefont{D.}~\bibnamefont{Bogorin}},
  \bibinfo{author}{\bibfnamefont{C.}~\bibnamefont{Cen}},
  \bibinfo{author}{\bibfnamefont{C.}~\bibnamefont{Bark}},
  \bibinfo{author}{\bibfnamefont{J.}~\bibnamefont{Park}},
  \bibinfo{author}{\bibfnamefont{C.}~\bibnamefont{Eom}}, \bibnamefont{and}
  \bibinfo{author}{\bibfnamefont{J.}~\bibnamefont{Levy}},
  \bibinfo{journal}{Appl. Phys. Lett.} \textbf{\bibinfo{volume}{97}},
  \bibinfo{pages}{173110} (\bibinfo{year}{2010}).

\bibitem[{\citenamefont{Kalinin et~al.}(2011)\citenamefont{Kalinin, Jesse,
  Tselev, Baddorf, and Balke}}]{Kalinin}
\bibinfo{author}{\bibfnamefont{S.}~\bibnamefont{Kalinin}},
  \bibinfo{author}{\bibfnamefont{S.}~\bibnamefont{Jesse}},
  \bibinfo{author}{\bibfnamefont{A.}~\bibnamefont{Tselev}},
  \bibinfo{author}{\bibfnamefont{A.}~\bibnamefont{Baddorf}}, \bibnamefont{and}
  \bibinfo{author}{\bibfnamefont{N.}~\bibnamefont{Balke}},
  \bibinfo{journal}{ACS nano} \textbf{\bibinfo{volume}{5}},
  \bibinfo{pages}{5683} (\bibinfo{year}{2011}).

\bibitem[{\citenamefont{Waser et~al.}(2009)\citenamefont{Waser, Dittmann,
  Staikov, and Szot}}]{Waser2009}
\bibinfo{author}{\bibfnamefont{R.}~\bibnamefont{Waser}},
  \bibinfo{author}{\bibfnamefont{R.}~\bibnamefont{Dittmann}},
  \bibinfo{author}{\bibfnamefont{G.}~\bibnamefont{Staikov}}, \bibnamefont{and}
  \bibinfo{author}{\bibfnamefont{K.}~\bibnamefont{Szot}},
  \bibinfo{journal}{Advanced Materials} \textbf{\bibinfo{volume}{21}},
  \bibinfo{pages}{2632} (\bibinfo{year}{2009}).

\bibitem[{\citenamefont{Stephenson and Highland}(2011)}]{Stephenson2011}
\bibinfo{author}{\bibfnamefont{G.~B.} \bibnamefont{Stephenson}}
  \bibnamefont{and} \bibinfo{author}{\bibfnamefont{M.~J.}
  \bibnamefont{Highland}}, \bibinfo{journal}{Phys. Rev. B}
  \textbf{\bibinfo{volume}{84}}, \bibinfo{pages}{064107}
  (\bibinfo{year}{2011}).

\bibitem[{\citenamefont{Morozovska et~al.}(2010)\citenamefont{Morozovska,
  Eliseev, Svechnikov, Krutov, Shur, Borisevich, Maksymovych, and
  Kalinin}}]{Morozovska2010}
\bibinfo{author}{\bibfnamefont{A.}~\bibnamefont{Morozovska}},
  \bibinfo{author}{\bibfnamefont{E.}~\bibnamefont{Eliseev}},
  \bibinfo{author}{\bibfnamefont{S.}~\bibnamefont{Svechnikov}},
  \bibinfo{author}{\bibfnamefont{A.}~\bibnamefont{Krutov}},
  \bibinfo{author}{\bibfnamefont{V.}~\bibnamefont{Shur}},
  \bibinfo{author}{\bibfnamefont{A.}~\bibnamefont{Borisevich}},
  \bibinfo{author}{\bibfnamefont{P.}~\bibnamefont{Maksymovych}},
  \bibnamefont{and} \bibinfo{author}{\bibfnamefont{S.}~\bibnamefont{Kalinin}},
  \bibinfo{journal}{Phys. Rev. B} \textbf{\bibinfo{volume}{81}},
  \bibinfo{pages}{205308} (\bibinfo{year}{2010}).

\bibitem[{\citenamefont{Kalinin and Bonnell}(2004)}]{Kalinin2004}
\bibinfo{author}{\bibfnamefont{S.}~\bibnamefont{Kalinin}} \bibnamefont{and}
  \bibinfo{author}{\bibfnamefont{D.}~\bibnamefont{Bonnell}},
  \bibinfo{journal}{Nano Lett.} \textbf{\bibinfo{volume}{4}},
  \bibinfo{pages}{555} (\bibinfo{year}{2004}).

\bibitem[{\citenamefont{Gruverman et~al.}(2009)\citenamefont{Gruverman, Wu, Lu,
  Wang, Jang, Folkman, Zhuravlev, Felker, Rzchowski, Eom
  et~al.}}]{Gruverman2009}
\bibinfo{author}{\bibfnamefont{A.}~\bibnamefont{Gruverman}},
  \bibinfo{author}{\bibfnamefont{D.}~\bibnamefont{Wu}},
  \bibinfo{author}{\bibfnamefont{H.}~\bibnamefont{Lu}},
  \bibinfo{author}{\bibfnamefont{Y.}~\bibnamefont{Wang}},
  \bibinfo{author}{\bibfnamefont{H.}~\bibnamefont{Jang}},
  \bibinfo{author}{\bibfnamefont{C.}~\bibnamefont{Folkman}},
  \bibinfo{author}{\bibfnamefont{M.}~\bibnamefont{Zhuravlev}},
  \bibinfo{author}{\bibfnamefont{D.}~\bibnamefont{Felker}},
  \bibinfo{author}{\bibfnamefont{M.}~\bibnamefont{Rzchowski}},
  \bibinfo{author}{\bibfnamefont{C.}~\bibnamefont{Eom}}, \bibnamefont{et~al.},
  \bibinfo{journal}{Nano Lett.} \textbf{\bibinfo{volume}{9}},
  \bibinfo{pages}{3539} (\bibinfo{year}{2009}).

\bibitem[{\citenamefont{Rondinelli et~al.}(2007)\citenamefont{Rondinelli,
  Stengel, and Spaldin}}]{Rondinelli2007}
\bibinfo{author}{\bibfnamefont{J.}~\bibnamefont{Rondinelli}},
  \bibinfo{author}{\bibfnamefont{M.}~\bibnamefont{Stengel}}, \bibnamefont{and}
  \bibinfo{author}{\bibfnamefont{N.}~\bibnamefont{Spaldin}},
  \bibinfo{journal}{Nature Nanotechnology} \textbf{\bibinfo{volume}{3}},
  \bibinfo{pages}{46} (\bibinfo{year}{2007}).

\bibitem[{\citenamefont{Burton and Tsymbal}(2009)}]{Burton2009}
\bibinfo{author}{\bibfnamefont{J.~D.} \bibnamefont{Burton}} \bibnamefont{and}
  \bibinfo{author}{\bibfnamefont{E.~Y.} \bibnamefont{Tsymbal}},
  \bibinfo{journal}{Phys. Rev. B} \textbf{\bibinfo{volume}{80}},
  \bibinfo{pages}{174406} (\bibinfo{year}{2009}).

\bibitem[{\citenamefont{Vaz et~al.}(2010)\citenamefont{Vaz, Hoffman, Segal,
  Reiner, Grober, Zhang, Ahn, and Walker}}]{Vaz2010}
\bibinfo{author}{\bibfnamefont{C.~A.~F.} \bibnamefont{Vaz}},
  \bibinfo{author}{\bibfnamefont{J.}~\bibnamefont{Hoffman}},
  \bibinfo{author}{\bibfnamefont{Y.}~\bibnamefont{Segal}},
  \bibinfo{author}{\bibfnamefont{J.~W.} \bibnamefont{Reiner}},
  \bibinfo{author}{\bibfnamefont{R.~D.} \bibnamefont{Grober}},
  \bibinfo{author}{\bibfnamefont{Z.}~\bibnamefont{Zhang}},
  \bibinfo{author}{\bibfnamefont{C.~H.} \bibnamefont{Ahn}}, \bibnamefont{and}
  \bibinfo{author}{\bibfnamefont{F.~J.} \bibnamefont{Walker}},
  \bibinfo{journal}{Phys. Rev. Lett.} \textbf{\bibinfo{volume}{104}},
  \bibinfo{pages}{127202} (\bibinfo{year}{2010}).

\bibitem[{\citenamefont{Molegraaf et~al.}(2009)\citenamefont{Molegraaf,
  Hoffman, Vaz, Gariglio, van~der Marel, Ahn, and Triscone}}]{Molegraaf2009}
\bibinfo{author}{\bibfnamefont{H.}~\bibnamefont{Molegraaf}},
  \bibinfo{author}{\bibfnamefont{J.}~\bibnamefont{Hoffman}},
  \bibinfo{author}{\bibfnamefont{C.}~\bibnamefont{Vaz}},
  \bibinfo{author}{\bibfnamefont{S.}~\bibnamefont{Gariglio}},
  \bibinfo{author}{\bibfnamefont{D.}~\bibnamefont{van~der Marel}},
  \bibinfo{author}{\bibfnamefont{C.}~\bibnamefont{Ahn}}, \bibnamefont{and}
  \bibinfo{author}{\bibfnamefont{J.}~\bibnamefont{Triscone}},
  \bibinfo{journal}{Advanced Materials} \textbf{\bibinfo{volume}{21}},
  \bibinfo{pages}{3470} (\bibinfo{year}{2009}).

\bibitem[{\citenamefont{Junquera and Ghosez}(2003)}]{Junquera2003a}
\bibinfo{author}{\bibfnamefont{J.}~\bibnamefont{Junquera}} \bibnamefont{and}
  \bibinfo{author}{\bibfnamefont{P.}~\bibnamefont{Ghosez}},
  \bibinfo{journal}{Nature} \textbf{\bibinfo{volume}{422}},
  \bibinfo{pages}{506} (\bibinfo{year}{2003}).

\bibitem[{\citenamefont{Zhuravlev et~al.}(2005)\citenamefont{Zhuravlev,
  Sabirianov, Jaswal, and Tsymbal}}]{Zhuravlev2005}
\bibinfo{author}{\bibfnamefont{M.}~\bibnamefont{Zhuravlev}},
  \bibinfo{author}{\bibfnamefont{R.}~\bibnamefont{Sabirianov}},
  \bibinfo{author}{\bibfnamefont{S.}~\bibnamefont{Jaswal}}, \bibnamefont{and}
  \bibinfo{author}{\bibfnamefont{E.}~\bibnamefont{Tsymbal}},
  \bibinfo{journal}{Phys. Rev. Lett.} \textbf{\bibinfo{volume}{94}},
  \bibinfo{pages}{246802} (\bibinfo{year}{2005}).

\bibitem[{\citenamefont{Kohlstedt et~al.}(2005)\citenamefont{Kohlstedt,
  Pertsev, Rodr{\'\i}guez~Contreras, and Waser}}]{Kohlstedt2005}
\bibinfo{author}{\bibfnamefont{H.}~\bibnamefont{Kohlstedt}},
  \bibinfo{author}{\bibfnamefont{N.}~\bibnamefont{Pertsev}},
  \bibinfo{author}{\bibfnamefont{J.}~\bibnamefont{Rodr{\'\i}guez~Contreras}},
  \bibnamefont{and} \bibinfo{author}{\bibfnamefont{R.}~\bibnamefont{Waser}},
  \bibinfo{journal}{Phys. Rev. B} \textbf{\bibinfo{volume}{72}},
  \bibinfo{pages}{125341} (\bibinfo{year}{2005}).

\bibitem[{\citenamefont{Wu et~al.}(2011)\citenamefont{Wu, Lee, Chen, Chang,
  Chen, Liang, Yu, He, Ramesh, and Chu}}]{Wu2011}
\bibinfo{author}{\bibfnamefont{C.-L.} \bibnamefont{Wu}},
  \bibinfo{author}{\bibfnamefont{P.-W.} \bibnamefont{Lee}},
  \bibinfo{author}{\bibfnamefont{Y.-C.} \bibnamefont{Chen}},
  \bibinfo{author}{\bibfnamefont{L.-Y.} \bibnamefont{Chang}},
  \bibinfo{author}{\bibfnamefont{C.-H.} \bibnamefont{Chen}},
  \bibinfo{author}{\bibfnamefont{C.-W.} \bibnamefont{Liang}},
  \bibinfo{author}{\bibfnamefont{P.}~\bibnamefont{Yu}},
  \bibinfo{author}{\bibfnamefont{Q.}~\bibnamefont{He}},
  \bibinfo{author}{\bibfnamefont{R.}~\bibnamefont{Ramesh}}, \bibnamefont{and}
  \bibinfo{author}{\bibfnamefont{Y.-H.} \bibnamefont{Chu}},
  \bibinfo{journal}{Phys. Rev. B} \textbf{\bibinfo{volume}{83}},
  \bibinfo{pages}{020103} (\bibinfo{year}{2011}).

\bibitem[{\citenamefont{Nonnenmann et~al.}(2010)\citenamefont{Nonnenmann,
  Gallo, and Spanier}}]{Nonnenmann2010}
\bibinfo{author}{\bibfnamefont{S.}~\bibnamefont{Nonnenmann}},
  \bibinfo{author}{\bibfnamefont{E.}~\bibnamefont{Gallo}}, \bibnamefont{and}
  \bibinfo{author}{\bibfnamefont{J.}~\bibnamefont{Spanier}},
  \bibinfo{journal}{Appl. Phys. Lett.} \textbf{\bibinfo{volume}{97}},
  \bibinfo{pages}{102904} (\bibinfo{year}{2010}).

\bibitem[{\citenamefont{Stengel et~al.}(2011)\citenamefont{Stengel,
  Aguado-Puente, Spaldin, and Junquera}}]{Stengel2011a}
\bibinfo{author}{\bibfnamefont{M.}~\bibnamefont{Stengel}},
  \bibinfo{author}{\bibfnamefont{P.}~\bibnamefont{Aguado-Puente}},
  \bibinfo{author}{\bibfnamefont{N.~A.} \bibnamefont{Spaldin}},
  \bibnamefont{and} \bibinfo{author}{\bibfnamefont{J.}~\bibnamefont{Junquera}},
  \bibinfo{journal}{Phys. Rev. B} \textbf{\bibinfo{volume}{83}},
  \bibinfo{pages}{235112} (\bibinfo{year}{2011}).

\bibitem[{\citenamefont{Stengel}(2011)}]{Stengel2011b}
\bibinfo{author}{\bibfnamefont{M.}~\bibnamefont{Stengel}},
  \bibinfo{journal}{Phys. Rev. B} \textbf{\bibinfo{volume}{84}},
  \bibinfo{pages}{205432} (\bibinfo{year}{2011}).

\bibitem[{\citenamefont{Schlom et~al.}(2007)\citenamefont{Schlom, Chen, Eom,
  Rabe, Streiffer, and Triscone}}]{Schlom2007}
\bibinfo{author}{\bibfnamefont{D.}~\bibnamefont{Schlom}},
  \bibinfo{author}{\bibfnamefont{L.}~\bibnamefont{Chen}},
  \bibinfo{author}{\bibfnamefont{C.}~\bibnamefont{Eom}},
  \bibinfo{author}{\bibfnamefont{K.}~\bibnamefont{Rabe}},
  \bibinfo{author}{\bibfnamefont{S.}~\bibnamefont{Streiffer}},
  \bibnamefont{and} \bibinfo{author}{\bibfnamefont{J.}~\bibnamefont{Triscone}},
  \bibinfo{journal}{Annu. Rev. Mater. Res.} \textbf{\bibinfo{volume}{37}},
  \bibinfo{pages}{589} (\bibinfo{year}{2007}).

\bibitem[{\citenamefont{Junquera et~al.}(2003)\citenamefont{Junquera, Zimmer,
  Ordejon, and Ghosez}}]{Junquera2003}
\bibinfo{author}{\bibfnamefont{J.}~\bibnamefont{Junquera}},
  \bibinfo{author}{\bibfnamefont{M.}~\bibnamefont{Zimmer}},
  \bibinfo{author}{\bibfnamefont{P.}~\bibnamefont{Ordejon}}, \bibnamefont{and}
  \bibinfo{author}{\bibfnamefont{P.}~\bibnamefont{Ghosez}},
  \bibinfo{journal}{Phys. Rev. B} \textbf{\bibinfo{volume}{67}}
  (\bibinfo{year}{2003}).

\end{thebibliography}

\end{document}